\documentclass[11pt]{article}
\usepackage[utf8]{inputenc}
\usepackage[english]{babel}
\usepackage{amsmath}
\usepackage{mathrsfs}
\usepackage{amssymb}
\usepackage{amsthm}
\usepackage{amsfonts}
\usepackage{xspace}
\usepackage[normalem]{ulem}
\usepackage{graphicx}
\usepackage[caption=false]{subfig} 
\usepackage{multirow}
\usepackage{appendix}
\usepackage{enumerate}
\usepackage{natbib}
\usepackage{url} 
\usepackage{authblk}

\usepackage{xcolor}
\usepackage{soul}

\newcommand{\blind}{1}

\newtheorem{theorem}{Theorem}
\newtheorem{corollary}{Corollary}

\newcommand{\bbeta}{\boldsymbol{\beta}}
\newcommand{\balpha}{\boldsymbol{\alpha}}
\newcommand{\btheta}{\boldsymbol{\theta}}
\newcommand{\bpsi}{\boldsymbol{\psi}}
\newcommand{\bX}{\boldsymbol{X}}
\newcommand{\bW}{\boldsymbol{W}}
\newcommand{\bZ}{\boldsymbol{Z}}
\newcommand{\indep}{\mbox{$\perp\!\!\!\perp$}}

\begin{document}

\def\spacingset#1{\renewcommand{\baselinestretch}%
{#1}\small\normalsize} \spacingset{1}

\title{\bf Model-Robust Inference for Clinical Trials that Improve Precision by Stratified Randomization and Covariate Adjustment}


\if1\blind
{
\author[1]{Bingkai Wang}
\author[2]{Ryoko Susukida}
\author[2]{Ramin Mojtabai}
\author[2,3]{Masoumeh Amin-Esmaeili}
\author[1]{Michael Rosenblum
    }

\affil[1]{\small Department of Biostatistics, Johns Hopkins Bloomberg School of Public Health, MD, USA}
\affil[2]{\small Department of Mental Health, Johns Hopkins Bloomberg School of Public Health, MD, USA}
\affil[3]{\small Iranian National Center for Addiction Studies (INCAS), Tehran University of Medical Sciences}
\date{\vspace{-5ex}}
\maketitle
} \fi

\if0\blind
{
\date{\vspace{-5ex}}
    \maketitle
} \fi

\begin{abstract}
Two commonly used methods for improving precision and power in clinical trials are stratified randomization and covariate adjustment. 
However, many trials do not 
fully capitalize on the combined precision gains from these two methods, which can lead to wasted resources in terms of sample size and trial duration. 
We derive consistency and asymptotic normality of model-robust estimators that combine these two methods, and show that these estimators can lead to substantial gains in precision and power. 
Our theorems cover a class of estimators that handle  continuous, binary, and time-to-event outcomes; missing outcomes under the missing at random assumption are handled as well.
For each estimator, we give a formula for a consistent variance estimator that is model-robust and that fully captures variance reductions from stratified randomization and covariate adjustment. 
Also, we give the first proof (to the best of our knowledge) of consistency and asymptotic normality of the Kaplan-Meier estimator under stratified randomization, and we derive its asymptotic variance.
The above results also hold for the biased-coin covariate-adaptive design. We demonstrate our results using three completed, phase 3, randomized  trial data sets of treatments for substance use disorder, where the variance reduction due to stratified randomization and covariate adjustment ranges from 1\% to 36\%. 
\end{abstract}

\noindent%

\spacingset{1.5} 
\section{Introduction}\label{sec:intro}
A joint guidance document from the U.S. Food and Drug Administration and the European Medicines Agency 
 \citep{ICH9} states that 
``Pretrial deliberations should identify those
covariates and factors expected to have an important influence on the primary variable(s),
and should consider how to account for these in the analysis to improve precision and to
compensate for any lack of balance between treatment groups.'' More recent regulatory guidance documents also encourage consideration of baseline variables in order to improve precision in randomized trials  \citep{EMAguideline2015,FDA2019,FDA2020}. 
Though there is a rich statistical literature on model-robust methods to adjust for baseline variables in randomized trials that use simple randomization,  less is known for trials that use other forms of randomization. This is a practical concern since, as discussed below, many clinical trials use other forms of randomization. 

``Covariate-adaptive randomization" refers to   randomization procedures that take baseline variables into account when assigning participants to study arms. The goal is to achieve better balance 
 across study arms in preselected strata of the baseline variables compared to simple randomization (which ignores baseline variables). E.g., balance on disease severity, a genetic marker, or another variable thought to be correlated with the primary outcome could be sought. The simplest  and most commonly used type of  covariate-adaptive randomization is stratified permuted block randomization  \citep{Zelen1974},  referred to as ``stratified randomization" throughout, for conciseness. 

 Compared with simple randomization, covariate-adaptive randomization can be advantageous in minimizing imbalance and improving efficiency \citep{Efron1971, Pocock&Simon1975, Wei1978}.
 Due to these benefits, covariate-adaptive randomization has become a popular approach in clinical trials. According to a survey by \cite{Lin2015}, 183 out of their sample of 224 randomized clinical trials published in 2014 in leading medical journals used some form of covariate-adaptive randomization. 
Stratified randomization was implemented by 70\% of trials in this survey. 
Another method for covariate-adaptive randomization is the biased-coin design by \cite{Efron1971}, which we call ``biased-coin randomization" throughout.
Other examples  include Wei's urn design \citep{Wei1978} and  rerandomization \citep{Lock2012}. We only consider the following two types of covariate-adaptive randomization: stratified randomization and    biased-coin randomization.

Concerns have been raised regarding how to perform valid statistical analyses at the end of trials that use   covariate-adaptive randomization. 
Adjusting for stratification variables  is recommended \citep{Lachin1988, Kahan2012, EMAguideline2015}. However, this recommendation is not reliably carried out. \cite{Kahan2012} sampled 65 published trials from major medical journals from March to May 2010 and found that 41 implemented covariate-adaptive randomization (among which 29 used stratified randomization), but only 14 adjusted in the primary analysis for the variables used in the randomization procedure. Furthermore, many  results 
on how to conduct the primary efficacy analysis in trials that use stratified randomization require one to assume a correctly specified regression model, e.g., \cite{Shao2010,Shao2013, Ma2015, Ma2018}. 
Our focus is on model-robust estimators, i.e., estimators that do not require such an assumption  when there is no missing data or when outcome data are missing completely at random; when censoring can depend on baseline variables, then additional assumptions are generally required.


\cite{YangTsiatis2001} showed that  the analysis of covariance (ANCOVA) estimator is consistent and asymptotically normal under simple randomization, and that this holds  under arbitrary misspecification of the linear regression model used to construct the estimator. Analogous results for the  ANCOVA estimator were shown by \cite{bugni2018}
under  a variety of covariate-adaptive  randomization procedures that include stratified and biased-coin randomization;
 however, their results 
 only allow  adjustment for the variables used  in the randomization procedure.
  The proofs of our results  build on key ideas from their work as  described below.
The results of \cite{Li2019} for the ANCOVA estimator are robust to arbitrary misspecification of the linear regression model; however, they use the randomization inference framework while 
many clinical trials are  analyzed using the superpopulation inference framework (as done here); see \cite{10.2307/3182793} for a comparison of these frameworks.
All of the results in this paragraph 
 are for the ANCOVA estimator, and so do not apply to logistic regression models for binary outcomes nor to commonly used models for time-to-event outcomes.
\cite{Ye2019} derived asymptotic distributions  for log-rank and score tests in survival analysis under covariate-adaptive randomization;
 however, estimation was not addressed. 


For trials using stratified or biased-coin randomization,  to the best of our knowledge, it was an open problem to determine (in the commonly used superpopulation inference framework and without making parametric model assumptions) the large sample properties of estimators
that involve any of the following features: binary or time-to-event outcomes, adjustment for baseline variables in addition to those in the randomization procedure, and missing data under  the missing at random assumption. This is the problem that we address, and we think that each of the above  features can be important in the analysis of clinical trials. For example,  binary and time-to-event outcomes  are commonly used in clinical trials. According to a survey by \cite{Austin2010} on trials published in leading medical journals in 2007, 74 out of  114 trials involved binary or time-to-event outcomes. As we show in our  data analyses, the addition of baseline variables beyond those used for stratified randomization can lead to substantial precision gains.
Handling missing data is also important  in order to avoid bias in treatment effect estimation. 

Under regularity conditions, we prove that a large class of estimators is consistent and  asymptotically normally distributed in randomized trials that use stratified or biased-coin randomization, and we give a formula for computing their asymptotic variance. This class of estimators consists of  all M-estimators \citep[p.41]{vaart_1998} that are consistent under simple randomization. It 
includes, e.g., the ANCOVA estimator for continuous outcomes; the standardized logistic regression estimator \citep{Scharfstein1999, Moore2009a} for binary outcomes; the doubly-robust weighted least squares (DR-WLS) estimator  from Marshall Joffe as described by \cite{robins2007} for continuous and binary outcomes; and the Kaplan-Meier (K-M) estimator \citep{Kaplan1958} of survival functions and the augmented inverse probability weighted (AIPW) estimator of the restricted mean survival time \citep{Diaz2019} for time-to-event outcomes. 

Our theorems imply that under standard regularity conditions,  whenever an estimator in our class is consistent and asymptotically normally distributed under simple randomization, then it is consistent and asymptotically normally distributed under stratified (or biased-coin) randomization with equal or smaller asymptotic variance. Also, its influence function is the same regardless of whether data is generated under simple, stratified or biased-coin randomization. This can be advantageous since for many estimators used to analyze randomized trials, their influence functions have already been derived under simple randomization. An estimator's influence function can be input into our formula (\ref{eq:est-var}) to produce a consistent variance estimator under stratified and biased-coin randomization. 


As in the aforementioned work, we assume that the randomization procedure and analysis method have been completely specified before the trial starts, as is typically required by regulators \citep{ICH9, EMAguideline2015, FDA2019, FDA2020}.


In the next section, we describe three trial examples to which we apply our methods. In Section~\ref{sec:def}, we describe our setup, notation and assumptions.  We present our main results in Section~\ref{sec:theory}. 
In Section~\ref{sec:estimators}, we give example estimators for continuous and binary outcomes to which our general results apply. In Section~\ref{sec:time-to-event}, we present asymptotic results for  the Kaplan-Meier estimator for  time-to-event outcomes.
Trial applications are provided in Section~\ref{sec:data-analysis}. Practical recommendations and future directions are discussed in Section~\ref{sec:discussion}.

\section{Three completed  trials that used stratified randomization}

In some cases below, the outcomes in our analyses differ from the primary outcomes in the corresponding trials. This is because we wanted to use similar outcomes across trials for illustration. Our outcomes are all considered  clinically important in the field of substance use disorder  treatments.

\subsection{Buprenorphine tapering and illicit opioid use  (NIDA-CTN-0003)}

The trial of ``Buprenorphine tapering schedule and illicit opioid use''  in the National Drug Abuse Treatment Clinical Trials Network (NIDA-CTN-0003), is a phase-3 randomized trial completed in 2005 \citep{CTN03}. The goal was to compare the effects of a short or long taper schedule after buprenorphine stabilization of patients with opioid use disorder. Patients were randomized into two arms: 28-day taper (control, 259 patients, 36\% missing outcomes) and 7-day taper (treatment, 252 patients, 21\% missing outcomes), stratified by maintenance dose (3 levels) measured at randomization. The outcome of interest is a binary indicator of whether a participant's urine tested at the end of the study is opioid-free. In addition to the stratification variable, we adjust for the following baseline variables: sex, opioid urine toxicology results, the Adjective Rating Scale for Withdrawal (ARSW), the Clinical Opiate Withdrawal Scale (COWS) and the Visual Analog Scale (VAS).

\subsection{Prescription opioid addiction treatment  (NIDA-CTN-0030)} 

The ``Prescription Opioid Addiction Treatment Study'' (NIDA-CTN-0030) is a phase-3 randomized trial completed in 2013 \citep{CTN30}. The goal was to determine whether adding individual drug counseling to the prescription of buprenorphine/naloxone would improve outcomes for patients with prescription opioid use disorder. 
Though this study adopted a 2-phase adaptive design, we focus on the first phase, in which  patients were randomized into standard medical management (control, 330 patients, 10\% missing outcomes) or standard medical management plus drug counseling (treatment, 335 patients, 13\% missing outcomes).
Randomization was stratified by the presence or absence of (i) a history of heroin use and (ii) current chronic pain, resulting in 4 strata. 
The outcome of interest is the proportion of negative urine laboratory results among all tests (treated as a continuous outcome between 0 and 1). Among all 5 urine laboratory tests during the first 4 weeks of phase I, if a patient missed two  consecutive visits,  then the outcome is regarded as missing. We included the following baseline variables in the analysis:  randomization stratum, age, sex and urine laboratory results.

\subsection{Internet-delivered treatment for substance abuse (NIDA-CTN-0044)}\label{sec:ctn44}

The  phase-3  randomized trial ``Internet-delivered treatment for substance abuse'' (NIDA-CTN-0044) was completed in 2012 \citep{CTN44}. The goal was to evaluate the effectiveness of a web-delivered behavioral intervention, Therapeutic Education
System (TES), in the treatment of substance abuse. Participants were randomly assigned to two arms: treatment as usual (control, 252 participants, 19\% missing outcomes) and treatment as usual plus TES (treatment, 255 participants, 18\% missing outcomes). Randomization was stratified by treatment site, patient's primary substance of abuse (stimulant or non-stimulant) and abstinence status at baseline. 
Since the available data set for this trial did not include the treatment site variable, only the patient's primary substance of abuse and abstinence status at baseline were included as strata (4 levels) in our analysis. 
After randomization, each participant was followed for 12 weeks with 2 urine laboratory tests per week. The outcome of interest is the proportion of negative urine lab results among all tests (treated as a continuous outcome between 0 and 1). If a participant missed visits of more than 6 weeks, the outcome is regarded as missing. We adjust for randomization stratum and the following additional baseline variables: age, sex and urine laboratory result. 

We also analyze a second outcome: time to abstinence, defined as the time to first two consecutive negative urine tests during the study. Censoring time is defined as the first missing visit. We used the data from the first 6 weeks of follow-up in our data analysis of this  time-to-event outcome, during which 99\% of the events occurred.

\section{Definitions and assumptions}\label{sec:def}
\subsection{Data generating distributions}
We focus on two-arm randomized trials that use simple, stratified or biased-coin randomization. Let $n$ denote the sample size.
For each participant $i = 1, \dots, n$, let $Y_i$ denote the primary outcome, $M_i$ denote whether $Y_i$ is non-missing ($M_i = 1$) or missing ($M_i = 0$), $A_i$ denote study arm assignment ($A_i=1$ if assigned to treatment and $A_i=0$ if assigned to control), and  $\bX_i$ denote a vector of baseline covariates.
This notation is for real-valued outcomes, e.g. continuous or binary outcomes. Modified definitions, assumptions, and results for time-to-event outcomes are in Section~\ref{sec:time-to-event}. 

We use the Neyman-Rubin potential outcomes framework \citep{neyman1990}, which  assumes the existence of potential outcomes $Y_i(0)$ and $Y_i(1)$ for each participant $i$. These represent the  outcome that would be observed under assignment to study arm $0$ or $1$, respectively. Though using potential outcomes introduces additional notation, it is needed in order to rigorously define the data generating distributions under the different randomization procedures that we consider.
We make the following consistency assumption linking the observed outcome $Y_i$ to the potential outcomes: 
$Y_i = Y_i(A_i) = Y_i(1)A_i + Y_i(0)(1-A_i)$ 
for each participant $i$.
Also, let $M_i(a)$ be the indicator of whether participant $i$ would have a non-missing outcome if they get assigned to study arm $a \in \{0,1\}$. We assume, analogous to the consistency assumption above, that  $M_i = M_i(A_i)=M_i(1)A_i + M_i(0)(1-A_i)$.

For each participant $i$, we define the full data vector (including potential outcomes, some of which are not observed) $\bW_i = (Y_i(1), Y_i(0), M_i(1), M_i(0), \bX_i)$  and the observed data vector 
 $\mathbf{O}_i=(A_i,\bX_i, Y_iM_i,M_i)$. The reason that the product $Y_iM_i$ appears in the observed data vector $\mathbf{O}_i$ is to encode that whenever the outcome is missing ($M_i=0$), the outcome value $Y_i$ is not available in $\mathbf{O}_i$ (since $Y_iM_i=0$). The data available to the analyst at the end of the trial are $\mathbf{O}_1,\dots,\mathbf{O}_n$. 
  
 We make the following assumptions on the distribution of $\{\bW_1,\dots,\bW_n\}$:

\noindent\textbf{Assumption 1. }

\begin{enumerate}
\item[(i)] $\bW_i, i = 1, \dots ,n$ are independent, identically distributed samples from an unknown  joint distribution $P$ on
$\bW = (Y(1), Y(0), M(1), M(0), \bX)$.
\item[(ii)] Missing at random: $M(a) \indep Y(a)|\bX$ for each arm $a \in \{0,1\}$,
where $\indep$ denotes independence.
\end{enumerate}

Throughout, we use $E$ to denote the expectation with respect to distribution $P$.

\subsection{Randomization procedures: simple, stratified, and biased-coin} \label{sec:randomization:procedures}
First consider simple randomization, which assigns study arms $A_1,\dots,A_n$ by independent Bernoulli draws each with fixed probability $\pi$ of being 1, e.g., using a random number generator. By design, the draws are independent of each other and of all   participant characteristics measured before randomization or not impacted by randomization. Therefore, we have that  $(A_1,\dots,A_n)$ is independent of $(\bW_1,\dots,\bW_n)$, and that the observed data  $\mathbf{O}_1,\dots,\mathbf{O}_n$ are independent, identically distributed.

Next consider stratified or biased-coin randomization, where treatment allocation depends on predefined baseline strata, such as gender, age, site, disease severity, or combinations of these. 
We refer to the baseline strata that are used in the randomization procedure as ``randomization strata". The baseline stratum of  participant $i$ is denoted 
 by the single, categorical variable $S_i$ taking $K$ possible values.
For example, if randomization strata are defined by 4 sites and a binary indicator of high disease severity, then $S$ has $K=8$  possible values. 
Let $S_i$ denote the stratification variable for participant $i$ and let $\mathcal{S} = \{1,\dots, K\}$ denote the set of all $K$ randomization strata. The goal of stratified or biased-coin randomization is to achieve balance in each stratum; that is, the proportion of participants assigned to the treatment arm is targeted to the prespecified proportion $\pi \in (0,1)$, e.g. $\pi = 0.5$. Throughout, 
the stratification variable $S$ is encoded in the  baseline covariate vector $\bX$ using $K-1$ dummy variables that make up the first $K-1$ components of $\bX$ (which can include additional baseline variables). 

Stratified randomization uses permuted blocks to assign treatment. For each randomization stratum, a randomly permuted block with fraction $\pi$ 1's (representing treatment) and $(1-\pi)$ 0's (representing control) is used for sequential allocation. When a  block is exhausted, a new block is used.

Biased-coin randomization can be applied when $\pi = 0.5$ and it allocates participants sequentially by  the following rule for $k = 1, \dots, n$:
\begin{displaymath}\arraycolsep=1.4pt\def\arraystretch{1.5}
P(A_k = 1|S_1, \dots, S_k, A_1, \dots, A_{k-1}) = \left\{\begin{array}{rc}
  0.5,   &  \mbox{ if } \sum_{i=1}^{k-1}(A_i - 0.5)I\{S_i = S_k\} = 0\\
 \lambda, & \mbox{ if }\sum_{i=1}^{k-1}(A_i - 0.5)I\{S_i = S_k\} < 0\\
 1-\lambda, & \mbox{ if }\sum_{i=1}^{k-1}(A_i - 0.5)I\{S_i = S_k\} > 0
\end{array}\right. 
\end{displaymath}
where $\lambda \in (0.5, 1]$, $I\{Z\}$ is the  indicator function that has value $1$ if $Z$ is true and $0$ otherwise, and by convention the first participant is assigned with probability $0.5$ to each arm. 
Our results for biased-coin randomization assume that   $\pi=0.5$. 

When comparing the three types of randomization procedures  (simple, stratified, or biased-coin), we assume that all use the same value of $\pi$.
For the stratified randomization and biased-coin  designs, it follows by construction (and was shown by \citealp{bugni2018})  that  
the study arm assignments $(A_1,\dots, A_n)$ are conditionally independent of the participant baseline variables and potential outcomes $(\bW_1,\dots,\bW_n)$ given the randomization strata $(S_1, \dots, S_n)$. Intuitively, this is because the study arm assignment procedure only has access to the participants' randomization  strata (and nothing else about the participants). 
Under stratified or biased-coin randomization, the observed data vectors   $\mathbf{O}_1,\dots,\mathbf{O}_n$ are not independent.

Under any of the three randomization procedures, 
 the observed data vectors $\mathbf{O}_1,\dots,\mathbf{O}_n$ are identically distributed; that is, the distribution of $\mathbf{O}_1$ is the same as that of $\mathbf{O}_2$, etc. Let $P^*$ denote 
 this distribution, i.e., the distribution of a generic, observed data vector $\mathbf{O}=(A,\bX, YM,M)$. This distribution is the same for each of the three randomization procedures, and 
 is that induced by first drawing a single realization $\bW=(Y(1), Y(0), M(1), M(0), \bX)$ from the distribution $P$ (see Assumption 1), then drawing $A$ as an independent Bernoulli draw with probability $\pi$ of being $1$, and lastly applying the consistency assumptions $Y = Y(1)A + Y(0)(1-A)$ and $M = M(1)A + M(0)(1-A)$ to construct  $Y$, the (non)-missingness indicator $M$, and their product $YM$. 
The corresponding expectation with respect to $P^*$ is denoted $E^*$, which is used below.
 The claims in this paragraph are proved in the Supplementary Material. 

\subsection{Targets of Inference (Estimands) and Estimators} \label{sec:M-estimator-defs}

For continuous and binary outcomes, our goal is to estimate a population parameter ${\Delta}^*$, which is a contrast between the marginal distributions of $Y(1)$ and $Y(0)$. For example, the parameter of interest can be the population average treatment effect, defined as $\Delta^*  = E[Y(1)] - E[Y(0)]$.

We consider  M-estimators of $\Delta^*$ \citep[Ch. 5]{vaart_1998}. 
Let $\btheta = (\Delta, \bbeta^t)^t$ denote a column vector of $p+1$ parameters where $\Delta \in \mathbb{R}$ is the parameter of interest and  $\bbeta \in \mathbb{R}^p$ is a column vector of $p$ nuisance parameters. We define the M-estimator 
$\widehat{\btheta}  = (\widehat{\Delta}, \widehat\bbeta^t)^t$ to be the solution to the following estimating  equations:
\begin{equation}\label{est-eq}
    \sum_{i=1}^n  \bpsi(A_i, \bX_i, Y_i, M_i; \btheta) = \boldsymbol{0},
\end{equation}
where $\bpsi$ is a column vector (with $p+1$ components) of known functions. We define $\widehat{\Delta}$ to be the estimator of $\Delta^*$.
We assume that $\bpsi(A,\bX,Y,M;\btheta)$ does not depend on the outcome $Y$ when $M=0$ (since $Y$ is missing in this case). 
Many estimators used in clinical trials, including all estimators defined in Section~\ref{subsec:examples}, can be expressed as solutions to estimating equations~(\ref{est-eq}) for  an appropriately chosen   estimating function $\bpsi$. 

For time-to-event outcomes, the K-M estimator of the survival curve is commonly used. Since it is not an M-estimator, our general result (Theorem~\ref{thm1}) for M-estimators below does not apply. We separately prove analogous results for the K-M estimator as described 
in Section~\ref{sec:time-to-event}.

We assume regularity conditions similar to the classical conditions that are used for proving consistency and asymptotic linearity of M-estimators for independent, identically distributed data, as given in Section 5.3 of \cite{vaart_1998}. One of the conditions is that the expectation of the estimating equations 
\begin{equation}
E^*[\bpsi(A, \bX, Y, M; \btheta)] = \boldsymbol{0}, \label{regCondition}
\end{equation}
has a unique solution in $\theta$, which is denoted as $\underline{\btheta} = (\underline{\Delta},\underline{\bbeta}^t)^t$.
The other regularity conditions are given in the Supplementary Material.  

We assume that the estimating equations 
  $\bpsi$ were chosen to ensure that the property  $\Delta^*=\underline{\Delta}$ holds. This property is generally needed to show consistency of the M-estimator $\widehat{\Delta}$ for $\Delta^*$ under simple randomization, and  
   has previously been proved for all of the estimators in Section~\ref{subsec:examples}. 
  In general, whether the property $\Delta^*=\underline{\Delta}$ holds does not depend on the randomization procedure (simple, stratified, or biased-coin randomization); this is because the property  depends only on $\bpsi$, $P$ and $P^*$. 

Results in Section~5.3 of \cite{vaart_1998} imply that under simple randomization, 
given Assumption 1 and the regularity conditions in the Supplementary Material,  $\widehat{\Delta}$ converges in probability to $\underline{\Delta}$ and is asymptotically normally distributed with asymptotic variance that we denote by $\widetilde{V}$. We focus on determining what happens under stratified or biased-coin randomization, where our main result (Section~\ref{sec:theory}) is that consistency and asymptotic normality still hold but   the asymptotic variance may be smaller (and a consistent variance estimator is given).

\section{Main result}\label{sec:theory}
Consider the setup in Section~\ref{sec:M-estimator-defs}, where 
 the M-estimator $\widehat{\Delta}$ is defined. We prove the following theorem: 

\begin{theorem}\label{thm1}
Assume the regularity conditions in the Supplementary Material,  $\Delta^*=\underline{\Delta}$, and Assumption 1. Under simple, stratified, or biased-coin 
randomization, we have  consistency ($\widehat{\Delta} \rightarrow \Delta^*$ in probability) and 
 \begin{equation}
    \sqrt{n}(\widehat{\Delta} - \Delta^*) = \frac{1}{\sqrt{n}}\sum_{i=1}^nIF(A_i, \bX_i, Y_i, M_i) + o_p(1), \label{asymptoticlinearity}
    \end{equation} 
	where the influence function $IF(A, \bX, Y, M)$ is the first entry of $-\boldsymbol{B}^{-1}\bpsi(A, \bX, Y, M; \underline{\btheta})$ for $\boldsymbol{B} = E^*\left[\frac{\partial}{\partial\btheta}\bpsi(A, \bX, Y, M; \btheta)\Big|_{\btheta = \underline{\btheta}}\right]$.

For stratified and biased-coin randomization, $\sqrt{n}(\widehat{\Delta} - \Delta^*) \xrightarrow{d} N(0,V)$ for 
\begin{equation}V = \widetilde{V} - \frac{1}{\pi(1-\pi)}E^*\left[E^*\left\{(A-\pi)IF(A, \bX, Y, M)|S\right\}^2\right], \label{variance:formula} \end{equation} 
	where $\widetilde{V} = E^*\{IF(A, \bX, Y, M)^2\}$ is the asymptotic variance under simple randomization. 
	The asymptotic variance $V$ can be consistently estimated by formula~(\ref{eq:est-var}) below. 
\end{theorem}

Theorem~\ref{thm1} implies that given our setup and assumptions, 
whenever an M-estimator $\widehat{\Delta}$ is consistent and asymptotically normally distributed under simple randomization, then it is consistent and asymptotically normally distributed under stratified (or biased-coin) randomization with equal or smaller asymptotic variance. Also, its influence function is the same regardless of whether data is generated under simple, stratified,   or biased-coin randomization; this can be advantageous since for many estimators used to analyze randomized trials, their influence functions have already been derived under simple randomization. The last display in Theorem~\ref{thm1} gives a simple formula for calculating the asymptotic variances  for these estimators under the other two randomization procedures, in terms of the influence function.

For the unadjusted estimator $\widehat{\Delta} = \sum_{i=1}^n Y_iA_i/\sum_{i=1}^n A_i - \sum_{i=1}^n Y_i(1-A_i)/\sum_{i=1}^n (1-A_i)$, our Theorem \ref{thm1} is equivalent to Theorem 4.1 of \cite{bugni2018} under stratified or biased-coin randomization. In the special case of continuous outcomes, if the ANCOVA estimator is used with $X = S$, then Theorem \ref{thm1} is equivalent to the result in section 4.2 of \cite{bugni2018} under stratified or biased-coin randomization, though their results also handle other types of covariate-adaptive randomization.

Theorem~\ref{thm1} above extends the results of \cite{bugni2018} to handle  the class of M-estimators, that is, estimators calculated by solving estimating equations~(\ref{est-eq}). This includes, for example, the ANCOVA estimator that adjusts for baseline covariates in addition to those used in the randomization procedure (Example 1 of Section~\ref{subsec:examples} below), the  standardized logistic regression estimator for binary outcomes (Example 2 of Section~\ref{subsec:examples}), and the DR-WLS estimator (Example 3 of Section~\ref{subsec:examples}).  This class of estimators also includes the  inverse-probability-weighted estimator (IPW, \citealp{Robins1994}), the  augmented inverse probability weighted estimator (AIPW, \citealp{Robins1994, Scharfstein1999}), the  Mixed-effects Model for Repeated Measures estimator (MMRM,  \citealp{doi:10.1081/BIP-120019265,doi:10.1080/10543400802609797, EMA-guideline2010}), and targeted maximum likelihood estimators (TMLE) that converge in 1-step \citep{Mark2012}, among others. Thus, Theorem~\ref{thm1} covers estimators that handle various outcome types, repeated measures outcomes, missing outcome data, and covariate adjustment.

We next give an overview of key ideas used in the proof of Theorem~\ref{thm1}. 
 The major challenge to prove asymptotic normality is that data from each participant are not independent of each other under the stratified or biased-coin randomization. 
The first steps are  to prove consistency and to show that  (\ref{asymptoticlinearity}) holds under each of the following data generating distributions:  simple, stratified, and biased-coin randomization. For simple randomization, these results follow directly from standard M-estimator theorems for independent, identically distributed data  \citep[Ch.5]{vaart_1998}, but  for the other randomization procedures new arguments are required.

     The second step is to  prove asymptotic normality and derive the formula for the asymptotic variance of the estimator under stratified and biased-coin randomization. This is more complicated than  under simple randomization 
     because the data from different participants are dependent. 
     Our proof relies on a generalization of  Lemmas B.1 and B.3 in the Supplement of \cite{bugni2018}, which is based on the empirical process result of \cite{shorack2009}.
     

In the third step, we prove consistency of  the following estimator for the asymptotic variance $V$, which is the following 
 empirical counterpart of the right side of  (\ref{variance:formula}): 
\begin{equation}\label{eq:est-var}
    \widehat{V} = \Tilde{V}_n - \frac{1}{\pi(1-\pi)} E_n\left[E_n\{(A-\pi)IF(A,\bX, Y,M)|S\}^2\right],
\end{equation} 
where $\Tilde{V}_n$ is the sandwich variance estimator of $\widehat{\Delta}$ (Section 3.2 of \citealp{tsiatis2007}), defined as the first-row first-column entry of
\begin{align*}
& \frac{1}{n}\left\{E_n\left[\frac{\partial}{\partial \btheta }\bpsi(A, \bX, Y, M; \btheta)\bigg|_{\btheta = \widehat{\btheta}}\right]\right\}^{-1}\left\{E_n\left[\bpsi(A, \bX, Y, M; \widehat\btheta)\bpsi(A, \bX, Y, M; \widehat\btheta)^t\right]\right\} \\
&\qquad \left\{E_n\left[\frac{\partial}{\partial \btheta }\bpsi(A, \bX, Y, M; \btheta)\bigg|_{\btheta = \widehat{\btheta}}\right]\right\}^{-1,t}
\end{align*}
and $E_n$ denotes expectation with respect to  the empirical distribution of the observed data $\textbf{O}_1,\dots,\textbf{O}_n$. 

\section{Examples of estimators for continuous and binary outcomes}\label{sec:estimators}

\subsection{ANCOVA, Standardized Logistic Regression, and DR-WLS Estimators}\label{subsec:examples}

We give several examples of estimators that our theorem above applies to. For estimators defined in Examples~1-3, the parameter of interest, i.e. $\Delta^*$, is the average treatment effect defined as $E[Y(1)]-E[Y(0)]$, and we denote $\bZ = (1,A, \bX^t)^t$. In Examples 1 and 2, we assume no missing data and we do not assume that the working models, i.e., the linear regression model in Example 1 and the logistic regression model in Example 2, are correctly specified. 


\vspace{5pt}
\noindent \textbf{Example 1.}
For continuous outcomes, the ANCOVA estimator $\widehat{\Delta}_{ancova}$ for $\Delta^*$ involves  first fitting a linear regression  working model 
$  E[Y|A, \bX] = \beta_0 + \Delta A + \bbeta_{\bX}^t\bX$ using ordinary least squares and then letting $\widehat{\Delta}_{ancova}$ be the estimate of $\Delta$. 
The ANCOVA estimator can be equivalently calculated by solving estimating equations (\ref{est-eq}) letting
\begin{displaymath}
    \bpsi(A, \bX, Y, M;\btheta) = 
        \{Y - (\beta_0 + \Delta A + \bbeta_{\bX}^t\bX)\}\bZ.
\end{displaymath}

\vspace{5pt}
\noindent \textbf{Example 2.}
For binary outcomes, the standardized logistic regression estimator $\widehat{\Delta}_{logistic}$ is calculated by first fitting a working model: $    P(Y=1|A,\bX) = \mbox{expit}(\beta_0 + \beta_AA + \bbeta_{\bX}^t\bX)$,
where $\mbox{expit}(x) = 1/(1+e^{-x})$, and getting the maximum likelihood estimates $(\widehat{\beta}_0, \widehat{\beta}_A, \widehat{\bbeta}_{\bX}^t)^t$. Then define
\begin{displaymath}
    \widehat{\Delta}_{logistic} = \frac{1}{n}\sum_{i=1}^n\{\mbox{expit}(\widehat\beta_0 + \widehat\beta_A + \widehat\bbeta_{\bX}^t\bX_i) - \mbox{expit}(\widehat\beta_0 + \widehat\bbeta_{\bX}^t\bX_i)\}.
\end{displaymath}
The estimator $\widehat{\Delta}_{logistic}$ can be equivalently calculated by solving estimating equations (\ref{est-eq}) letting
\begin{displaymath}
    \bpsi(A, \bX, Y, M;\btheta) = \left(\begin{array}{c}
        \mbox{expit}(\beta_0 + \beta_A + \bbeta_{\bX}^t\bX) -  \mbox{expit}(\beta_0 +  \bbeta_{\bX}^t\bX) - \Delta  \\
        \{Y - \mbox{expit}(\beta_0 + \beta_A A + \bbeta_{\bX}^t\bX)\}\bZ\\
    \end{array}\right).
\end{displaymath}
This estimator is mentioned as potentially useful in COVID-19 treatment and prevention trials in a recent FDA guidance \citep{FDA2020}.
Another estimator is the logistic coefficient estimator, defined as $\widehat\beta_A$. Unlike the standardized logistic regression estimator, $\widehat\beta_A$ estimates a conditional effect and can lead to invalid inference if there is treatment effect heterogeneity \citep{Steingrimsson2016}. We do not consider the logistic coefficient estimator below.

\vspace{5pt}
\noindent \textbf{Example 3.}
When some outcomes are missing, then one can estimate $\Delta^*$ by the DR-WLS estimator, which generalizes the estimators in Examples 1 and 2. However, due to missing data this estimator  requires additional assumptions (as is true for all estimators) described below.
The estimator is calculated by first fitting the logistic regression working model:
\begin{equation}\label{model:missingness}
P(M = 1|A, \bX) = \mbox{expit}(\alpha_0 + \alpha_A A + \balpha_{\bX}^t \bX)
\end{equation}
and getting the maximum likelihood estimates $(\widehat\alpha_0, \widehat\alpha_A, \widehat\balpha_{\bX}^t)^t$ of parameters $(\alpha_0, \alpha_A, \balpha_{\bX}^t)^t$. 
Next, fit the following working model for the outcome given study arm and baseline variables (from the generalized linear model family):
\begin{equation}\label{model:dr-wls}
E[Y|A,\bX] = g^{-1}(\beta_0 + \beta_AA + \bbeta_{\bX}^t\bX),
\end{equation}
with weights $1/\mbox{expit}(\widehat\alpha_0 + \widehat\alpha_A A_i + \widehat\balpha_{\bX}^t \bX_i)$ using only the data with $M_i = 1$. 
Here the inverse link function is $g^{-1}(x) = x$ for continuous outcomes and $g^{-1}(x) = \mbox{expit}(x)$ for binary outcomes.
Third, the DR-WLS estimator is
\begin{displaymath}
\widehat{\Delta}_{DR-WLS} = \frac{1}{n}\sum_{i =1}^n\{g^{-1}(\widehat\beta_0 + \widehat\beta_A + \widehat\bbeta_{\bX}^t\bX) - g^{-1}(\widehat\beta_0 + \widehat\bbeta_{\bX}^t\bX)\}.
\end{displaymath}
The DR-WLS estimator can be equivalently calculated by solving estimating equations (\ref{est-eq}) with
\begin{displaymath} \arraycolsep=1.4pt\def\arraystretch{1.5}
 \bpsi(A, \bX, Y, M;\btheta) = \left(\begin{array}{c}
      \{g^{-1}(\beta_0 + \beta_A + \bbeta_{\bX}^t \bX) - g^{-1}(\beta_0  + \bbeta_{\bX}^t \bX)\} - \Delta  \\
      \frac{M}{\mbox{expit}(\alpha_0 + \alpha_AA+ \balpha_{\bX}^t \bX)}\{Y - g^{-1}(\beta_0 + \beta_AA+ \bbeta_{\bX}^t \bX) \}\bZ \\
      \{M - \mbox{expit}(\alpha_0 + \alpha_AA+ \balpha_{\bX}^t \bX) \}\bZ
 \end{array}\right).
\end{displaymath}
For the DR-WLS estimator, we assume that at least one of the two working models (\ref{model:missingness}) and (\ref{model:dr-wls}) is correctly specified, and $ \inf_{(a, \mathbf{x}) \in (\mathcal{A}, \mathcal{X})}P(M = 1|a, \mathbf{x}) > 0$, where $(\mathcal{A}, \mathcal{X})$ is the support of $(A, \bX)$.
The ANCOVA estimator and the standardized logistic regression estimator are special cases of the DR-WLS estimator. If there are no missing data, which means $M_i = 1$ for $a = 0, 1$ and $i = 1, \dots, n$, and the regression weights used to fit (\ref{model:dr-wls}) are constant, then $\widehat{\Delta}_{DR-WLS}$ reduces to $\widehat{\Delta}_{ancova}$ for continuous outcomes and to $\widehat{\Delta}_{logistic}$ for binary outcomes. 
The DR-WLS estimator can be generalized to allow the addition of interaction terms in the model (\ref{model:dr-wls}).


\subsection{Asymptotic Results for Estimators in Examples 1-3}
Under simple randomization and assuming that 
 $\Delta^*=\underline{\Delta}$, consistency  and asymptotic normality for the estimators in Examples 1-3 have been proved by \cite{YangTsiatis2001, Scharfstein1999, robins2007}, respectively. Under stratified or biased-coin randomization, Theorem \ref{thm1} applies to these estimators since each is an M-estimator.
In particular, under the conditions in the theorem,  each of the three estimators is consistent  and asymptotically normal with 
 asymptotic  variance that is consistently estimated by (\ref{eq:est-var}). 

Under the additional conditions (a)-(c) listed in the corollary below, for each estimator in Examples 1-3, its asymptotic variance  is the same  regardless of whether simple, stratified, or  biased-coin randomization is used; also, the asymptotic variance is consistently estimated by the sandwich variance estimator $\widetilde{V}_n$.
We prove the corollary below by showing that each of the additional conditions (a)-(c) in Corollary~\ref{corollary:sandwich} implies that the rightmost term in equation~(\ref{variance:formula}) is 0. Under such conditions, the estimators and their corresponding sandwich variance estimators can be used to perform hypothesis tests and construct confidence intervals that are asymptotically correct. 

Recall that we assume throughout that $S$ is encoded by dummy variables in $\bX$.

\begin{corollary}\label{corollary:sandwich}
Assume that $\Delta^*=\underline{\Delta}$,  the regularity conditions in the Supplementary Material,  and Assumption 1. Consider the ANCOVA estimator or the standardized logistic regression estimator.
If any of the conditions (a)-(c) below holds, then  under simple, stratified, or biased-coin randomization, the estimator is consistent and asymptotically normally distributed with asymptotic variance $V = \widetilde{V}$; furthermore, the sandwich variance estimator is consistent. Conditions: 

\begin{itemize}
    \item[(a)] $\pi = 0.5$;
    \item[(b)] the outcome regression model (\ref{model:dr-wls}) includes indicators for the randomization strata and also  treatment-by-randomization-strata interaction terms;
    \item[(c)] the outcome regression model (\ref{model:dr-wls}) is correctly specified.
\end{itemize}
\end{corollary}
\noindent For the special case of the ANCOVA estimator with $\bX = S$, Corollary~\ref{corollary:sandwich} with condition~(a) or (b) was proved by \cite{bugni2018}. 
The claims in Corollary~\ref{corollary:sandwich} also hold for the 
the DR-WLS estimator if at least one of the two working models (\ref{model:missingness})  and (\ref{model:dr-wls}) is correctly specified and $ \inf_{(a, \mathbf{x}) \in (\mathcal{A}, \mathcal{X})}P(M = 1|a, \mathbf{x}) > 0$, where $(\mathcal{A}, \mathcal{X})$ is the support of $(A, \bX)$.

\section{Estimators involving time-to-event outcomes}\label{sec:time-to-event}


\subsection{Notation and Assumptions}
For time to event outcomes, we use slightly modified notation and assumptions compared to that above. We assume that the outcome is right-censored and that we observe either the failure (event) time or the censoring time, whichever occurs first. Let $Y_i$ denote the failure time and $M_i$ denote the censoring time. Other variables including $A_i, \bX_i$ and the potential outcomes $Y_i(a), M_i(a)$ for $a = 0,1$ are defined analogously as in Section~\ref{sec:def}. For each participant $i \in \{1,\dots, n\}$, we observe $(A_i, \bX_i, U_i, \delta_i)$, where $U_i = \min\{Y_i, M_i\}$ and $\delta_i = I\{Y_i \le M_i\}$. 
We further define a restriction time $\tau$ such that the time window $t\in [0,\tau]$ is of interest.
 We define $P^*$ and $E^*$ analogously as in Section~\ref{sec:randomization:procedures}, except here they represent the distribution and expectation, respectively, for a single observed data vector $(A, \bX, U,\delta)$.
 
The following assumption is made in place of Assumption 1:

\noindent\textbf{Assumption 1'.}
\begin{enumerate}
    \item [(i)] $\bW_i, i = 1, \dots ,n$ are independent, identically distributed samples from an unknown joint distribution $P$ on
$\bW = (Y(1), Y(0), M(1), M(0), \bX)$.

    \item [(ii)] Censoring completely at random: $M(a) \indep Y(a)$ for each arm $a \in \{0,1\}$.

    \item [(iii)] $P(\min\{Y(a), M(a)\} >\tau) > 0$ for each $a = 0, 1$.
\end{enumerate} 

Compared with Assumption 1, Assumption 1'(i) is the same as Assumption 1(i), and Assumption 1'(ii) assumes censoring completely at random instead of missing at random. This modification of the assumption on missing data is because we consider the K-M estimator and its consistency generally requires Assumption 1'(ii). Assumption 1'(iii) is often made in survival analysis, which states that there is a positive probability that both the failure time and censoring time occur after $\tau$ (under each study arm assignment). 

\subsection{Kaplan-Meier estimator under simple, stratified, and biased-coin randomization}
One commonly-used method for survival analysis is the K-M estimator. The goal is to estimate the survival curve $\{S_0^{(a)}(t): t\in[0,\tau]\}$ for each $a = 0,1$, where $S_0^{(a)}(t) = P(Y(a) > t)$. This represents the survival curve if everyone in the study population were assigned to study arm $a$.  The K-M estimator is defined as
\begin{displaymath}
    \widehat{S}_n^{(a)}(t) = \prod_{j: T_j  \le t}\left(1- \frac{\sum_{i=1}^n\delta_iI\{A_i =a\}I\{U_i = T_j\}}{\sum_{i=1}^nI\{A_i =a\}I\{U_i \ge T_j\}}\right),
\end{displaymath}
where $\{T_j, j= 1, \dots, m_n\}$ is the list of unique observed failure times. 

While the K-M estimator does not adjust for any baseline variable, its variance under simple randomization is typically different than under stratified or biased-coin randomization, and this is not accounted for by standard methods for estimating its variance. Specifically, the standard method for variance estimation will typically overestimate the K-M variance under stratified or biased-coin randomization, leading to wasted power. Our variance estimator below avoids this problem. Since the K-M estimator estimates a survival function rather than a real number or a vector, our Theorem~\ref{thm1} on M-estimators does not apply. The following theorem gives the asymptotic distribution of the K-M estimator under different types of randomization.

\begin{theorem}\label{thm2}
    Given Assumption 1', under simple, stratified, or biased-coin randomization, we have for each $t \in [0,\tau]$ that     \begin{equation}
        \sqrt{n}(\widehat{S}_n^{(a)}(t)-S_0^{(a)}(t)) = \frac{1}{\sqrt{n}}\sum_{i=1}^n IF^{(a)}(A_i, U_i, \delta_i; t) + o_{p^*}(1),
    \end{equation}
    where the influence function $IF^{(a)}(A_i, U_i, \delta_i; t)$ is defined in the Supplementary Material and $o_{p^*}(1)$ represents a sequence of random variables that converges to $0$ in probability uniformly over $t \in [0,\tau]$.
    
    For stratified and biased-coin randomization, the process \\ $\{\sqrt{n}(\widehat{S}_n^{(a)}(t)-S_0^{(a)}(t)):t\in[0,\tau]\}$ converges weakly to a mean $0$, tight Gaussian process with covariance function $V^{(a)}(t,t')$ defined in the Supplementary Material, which has the following property: for any $t \le \tau$, 
    \begin{displaymath}
       V^{(a)}(t,t) = \widetilde{V}^{(a)}(t,t) - \frac{1}{\pi(1-\pi)} E^*\left[E^*\left\{(A-\pi) IF^{(a)}(A, U, \delta;t)|S\right\}^2\right],
    \end{displaymath}
    where $\widetilde{V}^{(a)}(t,t)$ is the asymptotic variance under simple randomization. $V^{(a)}(t,t)$ can be consistently estimated as described in the Supplementary Material.
    
\end{theorem}

Analogous to Theorem~\ref{thm1}, Theorem~\ref{thm2} implies that the influence function of the K-M estimator is the same under simple, stratified, and  biased-coin  randomization.
Under simple randomization, the  influence function for the K-M estimator is given by  \cite{reid1981influence} for continuous survival functions and e.g., by \citet[Section 4.2]{kosorok2008} for survival functions that may have jumps. 

The above theorem implies that under stratified or biased-coin randomization, the K-M estimator is consistent and asymptotically normally distributed with equal or smaller asymptotic variance than under simple randomization. 
 The asymptotic covariance function of the K-M estimator under stratified or biased-coin randomization
 is given in Appendix C of the Supplementary Material. It can be used to construct pointwise confidence intervals and a simultaneous confidence band. 

The challenge in proving  Theorem~\ref{thm2} is that the traditional tool for deriving asymptotic normality in survival analysis, i.e., martingale central limit theorems  such as Theorem II.5.1 of \cite{andersen2012} or Theorem 5.1.1 of \cite{fleming2011counting}, is not applicable here because of the dependence among data points introduced by stratified or biased-coin randomization. 
To illustrate the intuition behind  this,  consider only the participants in one arm $a \in \{0, 1\}$.  For each participant $i$, define their observed failure counting process $N_i = \{N_i(t): t \in[0,\tau]\}$ for $N_i(t)=I\{U_i\le t, \delta_i = 1\}$
 and let $\Lambda_i(t)$ be the corresponding  cumulative hazard function. 
Under {\em simple randomization} and independent right censoring (Assumption 1'(ii)), the data from each participant are independent, identically distributed and 
the process $L_n = \{\sum_{i=1}^n N_i(t) - \sum_{i=1}^n\int_{t'=0}^tI\{Y_i\ge t'\}d\Lambda_i(t'): t\in[0,\tau]\}$ is a martingale given filtration $\mathcal{F}_t^{(n)} = \sigma\{N_i(s),I\{U_i \le s\}: s \in[0,t], i = 1,\dots, n\}$,  to which a martingale central limit theorem can be applied. However, when $N_1,\dots,N_n$ are {\em correlated},  $L_n$ may no longer be a martingale given filtration $\mathcal{F}_t^{(n)}$. To demonstrate this phenomenon, consider an extreme example with 
$n = 2$ participants, no censoring ($M_1=M_2=\infty$ with probability 1), $\tau = 0.5,$ and (correlated) event times for participants 1 and 2 having the joint distribution defined by $Y_1=- \log U$ and $Y_2= -\log (1-U)$ where $U$ is uniformly distributed on $[0,1]$. 
 Then $Y_1$ and $Y_2$ are dependent with each (marginally) having standard exponential distribution, i.e., $\Lambda_1(t) = \Lambda_2(t) = t$ for each $t \geq 0$. Direct calculation gives that  for any $0< t' < t \leq \tau$,
$E[L_2(t) -L_2(t') | \mathcal{F}_{t'}^{(2)}]$ equals $t'-t$ if $\min\{Y_1,Y_2\} \leq t'$ and equals $(t'-t)(2-2e^{-t'})/(1-2e^{-t'})$ otherwise. In either case, the conditional expectation is not equal to $0$ showing that 
$M_2$ is not a martingale. 

To overcome the above difficulty, in the proof of Theorem~\ref{thm2} we first developed a  central limit theorem for sums of random functions under stratified randomization (Lemma 4 in the Supplementary Material) based on the empirical process results of \cite{shorack2009} combined with generalizations of the techniques from  \cite{bugni2018}. We then proved Theorem~\ref{thm2} by generalizing the arguments in our proof of Theorem~\ref{thm1} to handle random functions. We conjecture that, using our central limit theorem, Theorem~\ref{thm2} can be generalized to apply to other estimators of survival functions, such as the covariate-adjusted estimators proposed by \cite{Lu2011, Zhang2015}, which may improve precision even further.

\subsection{Other estimators for time-to-event outcomes}
Another parameter of interest is the restricted mean survival time, defined as $\Delta^* = E[\min \{Y(1), \tau\}-\min \{Y(0), \tau\}]$. One covariate adjusted estimator of the restricted mean survival time is the augmented inverse probability weighted (AIPW) estimator of \cite{Moore2009}. This estimator is an M-estimator, to which our Theorem \ref{thm1} applies. 
When the survival probability at a given time point is the parameter of interest, one can use the K-M estimator or the method from \cite{Moore2009}.


\section{Clinical trial applications}\label{sec:data-analysis}
\subsection{Binary and continuous outcomes}
Table~\ref{tab:data-analysis} summarizes our data analyses involving binary and continuous outcomes. 
The outcome is binary for NIDA-CTN-0003 and is continuous for 
NIDA-CTN-0030 and NIDA-CTN-0044. In all cases, the target of inference is the average treatment effect defined as $E[Y(1)]-E[Y(0)]$.

All missing baseline values were imputed by the median for continuous variables and mode for binary or categorical variables. 
The only estimator in Table~\ref{tab:data-analysis} that adjusts for missing outcomes is 
the DR-WLS estimator; all other estimators omit data from the participants with missing outcomes.  
Negative (positive) estimates are in the direction of clinical benefit (harm). For all estimators presented in Table~\ref{tab:data-analysis}, the 95\% confidence interval (CI) is constructed using the normal approximation with variance calculated from formula~(\ref{eq:est-var}).

\begin{table}[htbp]
    \centering
    \caption{\small Summary of clinical trial data analyses with each cell giving the point estimate and 95\% CI of an estimator.  Each row is for a different estimator. 
    ``Adjusted estimator'' refers to the standardized logistic estimator for the trial with binary outcome (column 2) and to the ANCOVA estimator for the trials with continuous outcomes (columns 3 and 4).
    The variable in parentheses after the estimator name indicates which variables (if any) are adjusted for, with $S$ denoting the randomization strata only and $\bX$ denoting the randomization strata and additional baseline variables. 
}   
    \label{tab:data-analysis}
    \vspace{5pt}
    \resizebox{\textwidth}{!}{
    \begin{tabular}{lccc}
    &\multicolumn{3}{c}{Clinical Trial Identifier:} \\
    &  NIDA-CTN-0003 & NIDA-CTN-0030 & NIDA-CTN-0044\\
    \hline 
    Unadjusted  estimator & -0.104(-0.204, -0.004) & 0.015(-0.023, \ 0.052) & -0.093(-0.149, -0.038)\\ 
    Adjusted estimator ($S$) & -0.110(-0.209, -0.009)& 0.015(-0.022, \ 0.052) & -0.089(-0.145, -0.033)\\
    Adjusted estimator ($\bX$) & -0.104(-0.184, -0.024) & 0.012(-0.022, \ 0.046) & -0.087(-0.142, -0.032)\\
    DR-WLS estimator ($\bX$) & -0.099(-0.180, -0.019) & 0.012(-0.022, \ 0.045) & -0.091(-0.148, -0.035)\\
    \end{tabular}}
\end{table}

For NIDA-CTN-0003, the outcome is binary and  ``adjusted estimator" in Table~1 refers to the standardized logistic regression estimator. The unadjusted  point estimate was $-0.104$ with 95\%  CI $(-0.204, -0.004)$. If randomization strata and additional baseline variables are adjusted for (as in the row ``Adjusted estimator ($\bX$)" in Table~1), the point estimate is unchanged but the  95\% CI $(-0.184, -0.024)$ is substantially smaller. The corresponding variance reduction due to covariate adjustment, defined by one minus the variance ratio of ``Adjusted estimator ($\bX$)'' and the unadjusted estimator, is 36\%. This is equivalent to needing 36\% fewer participants to achieve the same power as a trial that uses the unadjusted estimator, asymptotically.

NIDA-CTN-0030 and NIDA-CTN-0044 had continuous-valued outcomes and ``Adjusted estimator" in Table~1 refers to the ANCOVA estimator. Covariate adjustment brings 17\% and 3\% variance reduction for NIDA-CTN-0030 and NIDA-CTN-0044, respectively, compared to the unadjusted estimator. When baseline variables are not strongly prognostic, such as in NIDA-CTN-0044, the variance reduction from additional baseline variables can be small.

In all three trials, the variance reduction due to adjusting for baseline variables beyond $S$, defined by one minus the variance ratio of ``adjusted estimator ($\bX$)'' and ``adjusted estimator ($S$)'', is the same (to the nearest percent) as the corresponding variance reduction  comparing 
``adjusted estimator ($\bX$)'' to the unadjusted estimator. 
This is expected for the ANCOVA estimator since \cite{bugni2018} showed that  ``adjusted estimator ($S$)'' and the unadjusted estimator are asymptotically equivalent when the randomization probability $\pi = 0.5$.
``DR-WLS estimator ($\bX$)'', which handles missing outcomes under the missing at random assumption, has a similar point estimate and 95\% CI compared to ``adjusted estimator ($\bX$)", which omits missing outcomes. 

For the unadjusted estimator, using the sandwich variance estimator instead of  formula~(\ref{eq:est-var}) may lead to conservative variance estimates, as implied by Theorem~\ref{thm1}. 
For example, for NIDA-CTN-0044, the 95\% CI of the unadjusted estimator constructed by formula~(\ref{eq:est-var}) is $(-0.149, -0.038)$, while the 95\% CI calculated using the sandwich variance formula is $(-0.162,-0.025)$, which is 23\% wider. 
The former 95\% CI is asymptotically correct assuming outcomes are missing completely at random. 
 Furthermore, the variance of the unadjusted estimator calculated by formula~(\ref{eq:est-var}) (which is consistent) is 34\% smaller than the variance calculated by the sandwich variance estimator (which is conservative).
In contrast, for the adjusted estimator or the DR-WLS estimator, since all three trials have randomization probabillity $\pi = 0.5$, the sandwich variance estimator is not conservative; this follows from Corollary~\ref{corollary:sandwich}. 


\subsection{Time-to-event outcome}
\begin{figure}[htb]
    \centering
    \includegraphics[width=\textwidth]{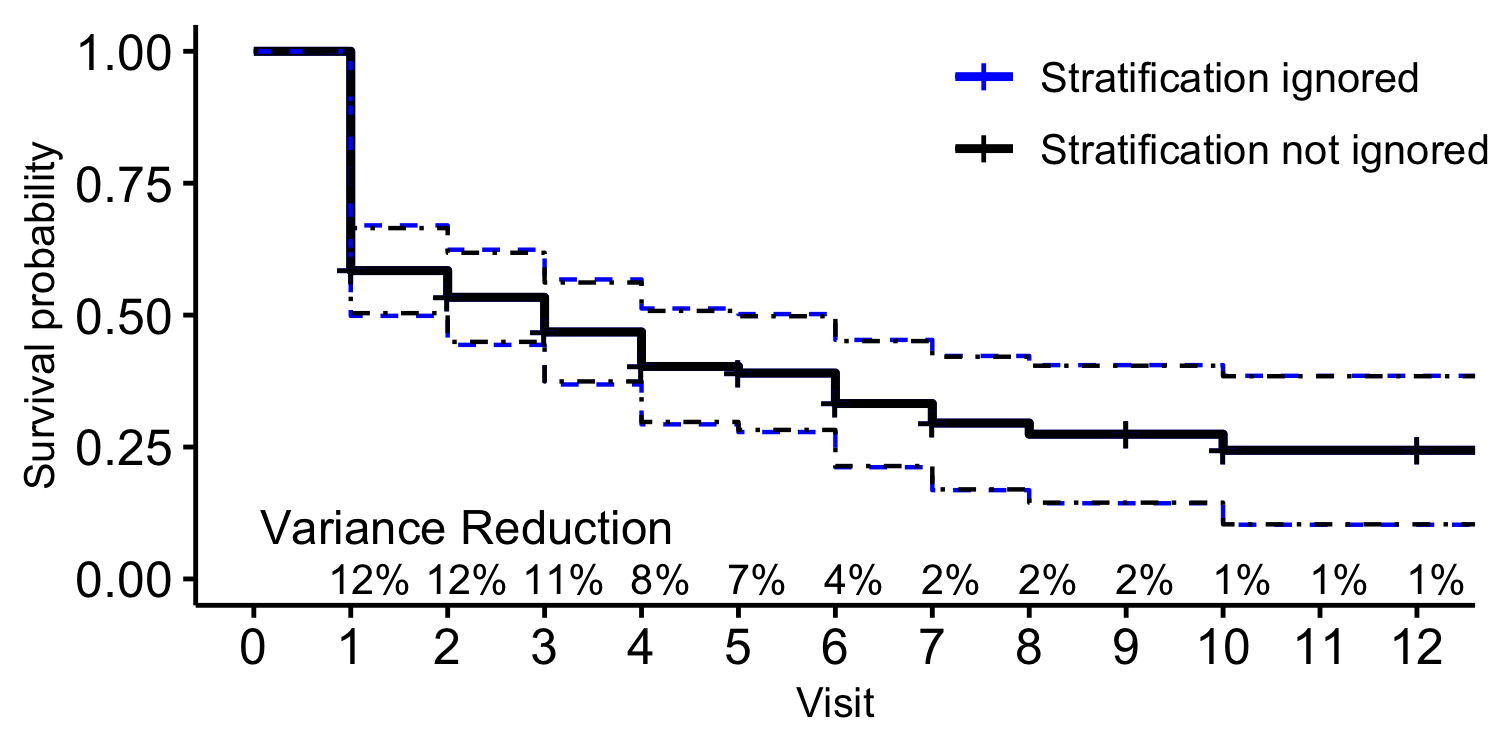}
    \caption{The K-M estimator of survival functions of NIDA-CTN-0044 treatment group. Dashed (dot-dashed) lines represent the estimates and confidence intervals if stratified randomization were (not) ignored in data analyses. ``Variance Reduction'' and the associated percentages represent the variance reduction by correcting the variance formula at each visit. The dashed and dot-dashed lines are very similar and almost coincide.}
    \label{fig:km}
\end{figure}

Figure~\ref{fig:km} presents the K-M estimator for time-to-abstinence in the treatment group as defined in Section~\ref{sec:ctn44} for study NIDA-CTN-0044. 
We estimated the variance of the K-M estimator in two different ways: one ignored the stratification variable and was the estimated variance returned by  the ``survfit'' function in R; the other  used our proposed variance formula that takes the stratification into account. For each of the two variance estimators, we constructed corresponding point-wise confidence intervals for the K-M estimator.

While Figure~\ref{fig:km} shows that confidence intervals based on different variance estimators are very close to each other, there are variance reductions due to accounting for stratification, which can be translated into sample size reduction needed to achieve the desired power. The variance reduction ranges from 1\% to 12\% as we consider the survival function at different time points. Among all time points, the first time point (one week after randomization) has the greatest variance reduction. Our findings are consistent with the result of Theorem~\ref{thm2}. For the control group, the results for the K-M estimator are given in the supplementary material, which are 
qualitatively similar to the treatment group. The variance formula from our theorem accounts for the improved precision due to stratified randomization (unlike standard methods that ignore stratification variables); this can be used to construct more powerful hypothesis tests based on the K-M estimator divided by its standard error.

\section{Discussion}\label{sec:discussion}


Our asymptotics, as in many asymptotic results under the commonly used superpopulation inference framework for randomized trials, assume  that the number of randomization strata is fixed and the number of participants in each stratum goes to infinity. This may be a reasonable approximation when no stratum has a small number of participants. In our data examples, the smallest stratum has 49 participants. An area of future research is to consider cases where some randomization strata have few participants.

In our data analysis of NIDA-CTN-0044, the stratification variable ``treatment site" was not  available in our data set. It was therefore neither used in the estimator nor in the corresponding variance estimate.
 The variance estimator (\ref{eq:est-var}) in this case is asymptotically conservative. This is because 
 the outer expectation in the rightmost term of (\ref{variance:formula}) is unchanged or decreased if $S$ is replaced by a coarsening of $S$ (defined as merging several randomization strata  together in a preplanned way, in the analysis); this follows from the conditional Jensen's inequality.
  This result may be useful more generally, e.g., when some strata are so small compared to the sample size that stratum-specific  evaluation of the empirical means $E_n$ in (\ref{eq:est-var}) cannot be reliably done. In such cases a pre-planned, coarsened stratum indicator could be used and the resulting hypothesis test would still control Type I error,  asymptotically.

Stratified randomization is related to stratified sampling designs, also called ``two-phase sampling''  \citep{SEN1988291,doi:10.1111/j.1467-9469.2006.00523.x, bai2013doubly}. 
To the best of our knowledge, asymptotic results for these designs do not directly apply to our problem; a key difference is that asymptotic results for stratified sampling designs often involve finite population inference (commonly used in survey sampling), while here we use superpopulation inference (commonly used in analyzing randomized trials). 

We provide R functions to calculate the variance for estimators including those in Examples 1-3 and the K-M estimator, which are available at \url{https://github.com/BingkaiWang/covariate-adaptive}.
 
\section*{Acknowledgment}
This project was supported by a research award from Arnold Ventures. The content is solely the responsibility of the authors and does not necessarily represent the official views of Arnold Ventures.
The information reported here results from secondary analyses of data from clinical trials conducted by the National Institute on Drug Abuse (NIDA). Specifically, data from NIDA–CTN-0003 (Suboxone (Buprenorphine/Naloxone) Taper: A Comparison of Two Schedules), NIDA-CTN-0030 (Prescription Opioid Addiction Treatment Study) and NIDA-CTN-0044 (Web-delivery of Evidence-Based, Psychosocial Treatment for Substance Use Disorders) were included. NIDA databases and information are available at (https://datashare.nida.nih.gov).

\bibliographystyle{apalike}
\bibliography{references}



\end{document}


\def\spacingset#1{\renewcommand{\baselinestretch}%
{#1}\small\normalsize} \spacingset{1}

\title{\bf Supplementary Material to ``Model-Robust Inference for Clinical Trials that Improve Precision by Stratified Randomization and Adjustment for Additional Baseline Variables''}


\if1\blind
{
 \author[1]{Bingkai Wang}
\author[2]{Ryoko Susukida}
\author[2]{Ramin Mojtabai}
\author[2,3]{Masoumeh Amin-Esmaeili}
\author[1]{Michael Rosenblum}

\affil[1]{\small Department of Biostatistics, Johns Hopkins Bloomberg School of Public Health, MD, USA}
\affil[2]{\small Department of Mental Health, Johns Hopkins Bloomberg School of Public Health, MD, USA}
\affil[3]{\small Iranian National Center for Addiction Studies (INCAS), Tehran University of Medical Sciences}
\date{\vspace{-5ex}}
\maketitle
} \fi

\if0\blind
{\date{\vspace{-5ex}}
    \maketitle
} \fi

\spacingset{1.5} 

In Section~\ref{app:regularity-conditions}, we give the regularity conditions in Theorem 1. In Section~\ref{app:variance-estimator}, we present consistent estimators of the asymptotic variance $V$ defined in Theorem 1. In Section~\ref{app:var-survivial}, we define the asymptotic variance $V^{(a)}(t,t')$, which is described in Theorem 2, and provide a consistent estimator for $V^{(a)}(t,t)$. In Section~\ref{app:proofs}, we give the proofs of Theorem 1 (Section~\ref{app:proofs-thm1}), Corollary 1 (Section~\ref{app:proofs-corollary1}) and Theorem 2 (Section~\ref{app:proofs-thm2}). In Section~\ref{sec:additional-analysis}, we give the results of the K-M estimator for the control group of NIDA-CTN-0044.

\section{Regularity conditions in Theorem~1}\label{app:regularity-conditions}
We denote the Euclidean norm as $||\cdot||$, which means $||\boldsymbol{x}|| = \sqrt{\boldsymbol{x}^t\boldsymbol{x}}$ for any $\boldsymbol{x}$ in a Euclidean space.
The regularity conditions for Theorem~1, which are  similar to those used in Section 5.3 of \cite{vaart_1998} in their theorem on  asymptotic normality of M-estimators (for independent, identically distributed data), are given below:

\hspace{5pt} (1) $\btheta \in \mathbf{\Theta}$, a compact set in $\mathbb{R}^{p+1}$.

\hspace{5pt} (2) $E[||\bpsi(a, \bX, Y(a), M(a); \btheta)||^2] < \infty$ for any $\btheta \in \mathbf{\Theta}$  and $a \in \{0, 1\}$.

\hspace{5pt} (3)
There exists a unique zero, denoted as $\underline{\btheta} = (\underline{\Delta},\underline{\beta}^t)^t$, of
\begin{equation*}
\pi E[\bpsi(1, \bX, Y(1), M(1); \btheta)] + (1-\pi) E[\bpsi(0, \bX, Y(0), M(0); \btheta)]=\boldsymbol{0}.
\end{equation*}

\hspace{5pt} (4) For each $a \in \{0,1\} $, the function $\btheta \mapsto \bpsi(a, x, y, m;\btheta)$ is  twice continuously  differentiable for every $(x, y, m)$ in the support of  $(\bX, Y(a), M(a))$ and is dominated by an integrable function $\boldsymbol{u}(\bX, Y(a), M(a))$.

\hspace{5pt} (5) There exist a $C > 0$ and and integrable function $v(\bX, Y(a), M(a))$, such that\\ $||\frac{\partial^2}{\partial\btheta\partial\btheta^t }\bpsi(a, x, y, m;\btheta)|| < v(x, y, m)$ element-wise for every $(a, x, y, m)$ in the support of $(A, \bX, Y(a), M(a))$  and $\btheta$ with $||\btheta - \underline{\btheta}|| < C$.

\hspace{5pt} (6) $E\left[\Big|\Big|\frac{\partial}{\partial\btheta}\bpsi(a, \bX, Y(a), M(a); \btheta)\Big|_{\btheta  =  \underline{\btheta}}\Big|\Big|^2\right] < \infty$ for $a \in \{0, 1\}$ and 
\begin{displaymath}
    \pi E\left[\frac{\partial}{\partial\btheta}\bpsi(1, \bX, Y(1), M(1); \btheta)\Big|_{\btheta  =  \underline{\btheta}}\right] + (1-\pi) E\left[\frac{\partial}{\partial\btheta}\bpsi(0, \bX, Y(0), M(0); \btheta)\Big|_{\btheta  =  \underline{\btheta}}\right]
\end{displaymath}
is invertible.

The above regularity conditions can be equivalently expressed in terms of distribution $P^*$ instead of $P$ (using the result of Lemma~\ref{lemma:E*} in Section D). 
Specifically, conditions (2), (3) and (6) are equivalent to conditions (2'), (3') and (6') below  respectively:

\hspace{5pt} (2') $E^*[||\bpsi(A, \bX, Y, M; \btheta)||^2] < \infty$ for any $\btheta \in \mathbf{\Theta}$.

\hspace{5pt} (3')
There exists a unique zero, denoted as $\underline{\btheta} = (\underline{\Delta},\underline{\beta}^t)^t$, of $E^*[\bpsi(A, \bX, Y, M; \btheta)]=\boldsymbol{0}$.

\hspace{5pt} (6') $E^*\left[\Big|\Big|\frac{\partial}{\partial\btheta}\bpsi(A, \bX, Y, M; \btheta)\Big|_{\btheta  =  \underline{\btheta}}\Big|\Big|^2\right] < \infty$ and $E^*\left[\frac{\partial}{\partial\btheta}\bpsi(A, \bX, Y, M; \btheta)\Big|_{\btheta  =  \underline{\btheta}}\right]$ is invertible.

\section{Consistent estimator of $V$ in Theorem 1}\label{app:variance-estimator}

We define
\begin{displaymath}
 \widehat{V} = \widetilde{V}_n - \frac{1}{\pi(1-\pi)}\sum_{s \in \mathcal{S}} \widehat{p}(s)\widehat{d}(s)^2,
\end{displaymath}
where $\widetilde{V}_n = \boldsymbol{e}_1^t \widehat{\boldsymbol{B}}^{-1} \widehat{\boldsymbol{C}} \widehat{\boldsymbol{B}}^{-1,t} \boldsymbol{e}_1$, $ \widehat{p}(s) = n^{-1}\sum_{i=1}^nI\{S_i = s\}$ and
\begin{align*}
    \widehat{d}(s) &= \frac{1}{n}\sum_{i=1}^n\frac{I\{S_i = s\}}{\widehat{p}(s)}\boldsymbol{e}_1^t\widehat{\boldsymbol{B}}^{-1}(A_i-\pi)\bpsi(A_i, \bX_i, Y_i, M_i; \widehat\btheta),
\end{align*}
where 
\begin{align*}
    \widehat{\boldsymbol{B}} &= \frac{1}{n}\sum_{i=1}^n\frac{\partial}{\partial\btheta} \bpsi(A_i, \bX_i, Y_i, M_i; \btheta) \bigg|_{\btheta = \widehat{\btheta}}, \\
     \widehat{\boldsymbol{C}} &= \frac{1}{n}\sum_{i=1}^n \bpsi(A_i, \bX_i, Y_i, M_i; \widehat\btheta)\bpsi(A_i, \bX_i, Y_i, M_i; \widehat\btheta)^t,
\end{align*}
$I\{Z\}$ is an indicator function which equals 1 if $Z$ is true and 0 otherwise, and $\boldsymbol{e}_1$ is a $(p+1)$-dimensional column vector with the first entry 1 and the rest 0. If $\widehat{p}(\widetilde{s}) = 0$ for some $\widetilde{s} \in \mathcal{S}$, which means there are no participants with $S_i =\widetilde{s}$, we let $d(\widetilde{s}) = 0$ by convention.

Consistency of $\widehat{V}$ to $V$ is shown in Section~\ref{app:proofs-thm1}.

\section{Expression of $V^{(a)}(t,t')$ and consistent estimator of $V^{(a)}(t,t)$ in Theorem 2}\label{app:var-survivial}
In the setting of survival analysis, we introduce the following notations in addition to those already defined in Section 6 of the main paper.

Define the potential survival function $S_0^{(a)}(t) = P(Y(a) > t)$ for each $a = 0,1$ and $t \in [0, \tau]$. Let $U(a) = \min\{Y(a), M(a)\}$ and $\delta(a) = I\{Y(a)\le M(a)\}$ for each $a \in \{0,1\}$. Define $N(t) = I\{U \le t, \delta = 1\}$ and $ N^{(a)}(t) = I\{U(a) \le t, \delta(a) = 1\}$ for each $a = 0, 1$ and $t \in [0, \tau]$. For each $a = 0, 1$ and $t \in [0, \tau]$, we further define 
\begin{equation}\label{KM-martingale}
H^{(a)}(t) = \int_{t' = 0}^{t' = t}\frac{d L^{(a)}(t')}{(1-\Delta\Lambda^{(a)}(t'))P(U(a)\ge t')},
\end{equation}
where
\begin{align*}
    L^{(a)}(t) &= N^{(a)}(t) - \int_{t' = 0}^{t' = t} I\{U(a)\ge t'\}d\Lambda^{(a)}(t'), \\
    \Lambda^{(a)}(t) &= E\left[\int_{t' = 0}^{t' = t} \frac{dN^{(a)}(t')}{P(U(a)\ge t')}\right],\\
    \Delta\Lambda^{(a)}(t) &= \Lambda^{(a)}(t) - \Lambda^{(a)}(t-),
\end{align*}
with $\Lambda^{(a)}(t-)$ representing the left limit of $\Lambda^{(a)}(t)$. Then $IF^{(a)}(A,U,\delta;t)$ is defined as
\begin{displaymath}
 IF^{(a)}(A,U,\delta;t) = - \frac{I\{A = a\}}{\pi_a}S_0^{(a)}(t)H^{(a)}(t).
\end{displaymath}

For $t, t' \in [0, \tau]$, we define
\begin{align*}
     V^{(a)}(t,t') &= S_0^{(a)}(t)S_0^{(a)}(t')\left\{\frac{1}{\pi_a}E[Cov(H^{(a)}(t), H^{(a)}(t')|S)]\right. \\
     &\qquad + Cov(E[H^{(a)}(t)|S], E[H^{(a)}(t')|S])\bigg\}, \numberthis \label{def-V}
\end{align*}
where $\pi_a = a\pi + (1-a)(1-\pi)$ for $a \in \{0,1\}$.
We estimate $V^{(a)}(t,t)$ by
\begin{equation}\label{hat_Vtt}
 \widehat{V}^{(a)}(t,t) = \frac{\widehat{S}_n^{(a)}(t)^2}{\pi_a} \left[\widehat{B}^{(a)}(t) - \frac{(1-\pi_a)}{\pi_a}\sum_{s \in \mathcal{S}}\widehat{p}(s) \left(\frac{\sum_{i=1}^nI\{S_i = s\}\widehat{H}^{(a)}_i(t)}{\sum_{i=1}^nI\{S_i = s\}} \right)^2\right],
\end{equation}
where $\widehat{S}_n^{(a)}(t)$ is the Kaplan-Meier estimator defined in Section 6 of the main paper,
\begin{align*}
    \widehat{B}^{(a)}(t) &= \int_{t'=0}^{t'=t}\frac{d\widehat{\Lambda}^{(a)}(t')}{\widehat{P}(U(a)\ge t')(1 - \Delta\widehat{\Lambda}^{(a)}(t'))},\\
    \widehat{H}^{(a)}_i(t) &= \int_{t'=0}^{t'=t}\frac{I\{A_i = a\}(dN_i(t') - I\{U_i \ge t'\}d\widehat{\Lambda}^{(a)}(t'))}{\pi_a\widehat{P}(U(a)\ge t')(1 - \Delta\widehat{\Lambda}^{(a)}(t'))},
\end{align*}
with
\begin{align*}
    \widehat{P}(U(a)\ge t) &= \frac{\sum_{i=1}^nI\{A_i = a\}I\{U_i \ge t\}}{\sum_{i=1}^nI\{A_i = a\}}, \\
    \widehat{\Lambda}^{(a)}(t) &= \int_{t'=0}^{t'=t} \frac{\sum_{i=1}^n I\{A_i = a\}dN_i(t')}{\sum_{i=1}^n I\{A_i = a\}I\{U_i \ge t'\}}, \\
    \Delta\widehat{\Lambda}^{(a)}(t) &= \frac{\sum_{i=1}^n I\{A_i = a\}I\{U_i = t, \delta_i = 1\}}{\sum_{i=1}^n I\{A_i = a\}I\{U_i \ge t\}}.
\end{align*}
Consistency of $\widehat{V}^{(a)}(t,t)$ to $V^{(a)}(t,t)$ is shown in Section~\ref{app:proofs-thm2}.

\section{Proofs} \label{app:proofs}
\begin{definition}\label{def:conditional-process}
Consider any random vector $\boldsymbol{U}$ taking values in $\mathbb{R}^m$ and any random variable $S$ taking  values  on  the  discrete  set $\mathcal{S}=\{1,...,K\}$. We assume that $\boldsymbol{U}$ and $S$ are defined on the same probability space and that $P(S=s)>0$ for each $s \in \mathcal{S}$.
We use the notation $\boldsymbol{U}^{(1)}, \dots, \boldsymbol{U}^{(K)}$ to represent the ordered list of $K$ independent random vectors where for each $s = 1, \dots, K$, the marginal distribution of $\boldsymbol{U}^{(s)}$ is the conditional distribution of $\boldsymbol{U}$ given $S=s$. Then for each $s \in \mathcal{S}$, $\boldsymbol{U}^{(s)}$ is called the conditional variable of $\boldsymbol{U}$ given $S=s$.
  
Next, consider any measurable function $\bh: \mathbb{R}^m \times \mathbf{\Theta} \rightarrow \mathbb{R}^q$ where $\mathbf{\Theta}$ is called the index set. Define 
the 
stochastic process $Z = \{\bh(\boldsymbol{U};\btheta):  \btheta \in \mathbf{\Theta}\}$.
Let $Z^{(1)},\dots,Z^{(K)}$ denote the ordered list of $K$ independent stochastic processes  where 
for each $s \in \mathcal{S}$ we define  
$Z^{(s)} = \{\bh(\boldsymbol{U}^{(s)};\btheta):  \btheta \in \mathbf{\Theta}\}$. Then for each $s \in \mathcal{S}$, $Z^{(s)}$ is called the conditional process of $Z$ given $S=s$, and $Z^{(s)}(\btheta)$ denotes the realization of $Z^{(s)}$ at $\btheta$, i.e., $Z^{(s)}(\btheta)=\bh(\boldsymbol{U}^{(s)}; \btheta)$.
\end{definition}

\subsection{Proof of Theorem 1}\label{app:proofs-thm1}
In this section, we first present four lemmas that are critical for proving our main results. Lemmas~\ref{lemma1} and \ref{lemma2} generalize results of \cite{bugni2018} and the proofs of them are very similar to \cite{bugni2018}. Lemmas~\ref{lemma:E*} connects $P$, which involves potential outcomes, and $P^*$, the distribution of observed data. Lemma~\ref{lemma:observed-data-distribution} proves the the claims in Section 3.2 of the main paper regarding $P^*$.
Then we prove Theorem~1 based on these lemmas.
\begin{lemma}\label{lemma1}
Let $Z(\btheta) = h(Y(1), Y(0), M(1), M(0), \bX;\btheta)$ for some measurable real-valued function $h$ and $\btheta \in \mathbf{\Theta}$. For each $s\in \mathcal{S}$, let $Z^{(s)}$ denote the conditional process of $ \{Z(\btheta): \btheta \in \mathbf{\Theta}\}$ given $S=s$ (using Definition~\ref{def:conditional-process}).  Supposing $Z_1(\btheta), \dots, Z_n(\btheta)$ are independent, identically distributed samples from the distribution of $Z(\btheta)$,
then under stratified randomization or the biased-coin design, \\
(1) if $E[|Z(\btheta)|] < \infty$ for each $\btheta \in \mathbf{\Theta}$, then  $\frac{1}{n}
\sum_{i=1}^nA_iZ_i(\btheta) \xrightarrow{P} \pi E[Z(\btheta)]$; \\
(2) if $\sup_{\btheta \in \mathbf{\Theta}}| E[Z^{(s)}(\btheta)]| < \infty$ and $Z^{(s)}$ is P-Glivenko-Cantelli for each $s \in \mathcal{S}$, then
$\sup_{\btheta \in \mathbf{\Theta}}\big|\frac{1}{n}\sum_{i=1}^n A_iZ_i(\btheta) - \pi E[Z(\btheta)]\big| \xrightarrow{P} 0 $ .
\end{lemma}
\begin{proof}
For the first part of the Lemma, see Lemma B.3 in the supplementary material of \cite{bugni2018}. The only difference is that we replace $(Y_i(1), Y_i(0),S_i)$ by $(Y_i(1), Y_i(0), M_i(1), M_i(0), \bX_i)$ and all of the arguments still hold.

For the second part of the Lemma, we give the proof that generalizes the arguments in Lemma B.3 in the supplementary material of \cite{bugni2018} to the supremum over a class of functions.
Under stratified randomization or the biased-coin design, we have $(Z_1(\btheta),\dots, Z_n(\btheta)) \indep (A_1, \dots, A_n) | (S_1,\dots, S_n)$. Then, for each $\btheta \in \mathbf{\Theta}$,  $\frac{1}{n}\sum_{i=1}^nA_iZ_i(\btheta)$ has the same distribution with the same quantity where $Z_i(\btheta)$ are ordered by strata and then by treatment group ($A_i=1$ first) within each stratum.
Independently for each $s \in \mathcal{S}$ and independently of $(A_1, \dots, A_n)$ and $(S_1, \dots, S_n)$, let $Z_1^{(s)}(\btheta), \dots, Z_n^{(s)}(\btheta)$ be independent, identically distributed samples from the distribution $Z^{(s)}(\btheta)$.  By construction, $\{\frac{1}{n}\sum_{i=1}^nA_iZ_i(\btheta)|(A_1,S_1,\dots, A_n,S_n)\}$ has the same distribution with \\ $\{\sum_{s\in\mathcal{S}}\frac{1}{n}\sum_{i=1}^{n_1(s)}Z_i^{(s)}(\btheta)|(A_1,S_1,\dots, A_n,S_n)\}$ , where $n_1(s) = \sum_{i=1}^n A_iI\{S_i=s\}$.
Hence $\frac{1}{n}\sum_{i=1}^nA_iZ_i(\btheta)$ has the same distribution with $\sum_{s\in\mathcal{S}}\frac{1}{n}\sum_{i=1}^{n_1(s)}Z_i^{(s)}(\btheta)$. Then it suffices to show $\sup_{\btheta \in \mathbf{\Theta}}|\frac{1}{n}\sum_{i=1}^{n_1(s)}Z_i^{(s)}(\btheta) -\pi P(S=s) E[Z^{(s)}(\btheta)]| \xrightarrow{P} 0$ for each $s \in \mathcal{S}$. 

Under stratified randomization or the biased-coin design, we have $\frac{n_1(s)}{n} \xrightarrow{P} \pi P(S=s)$. Since $n_1(s)$ does not involve $\btheta$, we have
\begin{align*}
    &\sup_{\btheta \in \mathbf{\Theta}}\bigg|\frac{1}{n}\sum_{i=1}^{n_1(s)}Z_i^{(s)}(\btheta) -\pi P(S=s) E[Z^{(s)}(\btheta)]\bigg| \\
    &= \frac{n_1(s)}{n}\sup_{\btheta \in \mathbf{\Theta}}\bigg|\frac{1}{n_1(s)}\sum_{i=1}^{n_1(s)}Z_i^{(s)}(\btheta) -\frac{n\pi P(S=s)}{n_1(s)} E[Z^{(s)}(\btheta)]\bigg| \\
    &\le  \frac{n_1(s)}{n}\sup_{\btheta \in \mathbf{\Theta}}\bigg|\frac{1}{n_1(s)}\sum_{i=1}^{n_1(s)}Z_i^{(s)}(\btheta) - E[Z^{(s)}(\btheta)]\bigg| + \frac{n_1(s)}{n}\sup_{\btheta \in \mathbf{\Theta}}\bigg|\left(\frac{n\pi P(S=s)}{n_1(s)}-1\right) E[Z^{(s)}(\btheta)]\bigg| \\
    &= (\pi P(S=s) + o_p(1))\sup_{\btheta \in \mathbf{\Theta}}\bigg|\frac{1}{n_1(s)}\sum_{i=1}^{n_1(s)}Z_i^{(s)}(\btheta) - E[Z^{(s)}(\btheta)]\bigg| + o_p(1)\sup_{\btheta \in \mathbf{\Theta}}| E[Z^{(s)}(\btheta)]|.
\end{align*}
Since $\sup_{\btheta \in \mathbf{\Theta}}| E[Z^{(s)}(\btheta)]| < \infty$, then it suffices to show $\sup_{\btheta \in \mathbf{\Theta}}|\frac{1}{n_1(s)}\sum_{i=1}^{n_1(s)}Z_i^{(s)}(\btheta) -\ E[Z^{(s)}(\btheta)]| \xrightarrow{P} 0$. To this end, by the almost sure representation theorem, we can construct $\frac{\Tilde{n}_1(s)}{n}$ such that $\frac{\Tilde{n}_1(s)}{n}$ has the same distribution with $\frac{n_1(s)}{n}$ and $\frac{\Tilde{n}_1(s)}{n} \rightarrow \pi P(S=s)$ almost surely. Since $(A_1, \dots, A_n)$ and $(S_1, \dots, S_n)$ are independent of $\{Z_1^{(s)}(\btheta), \dots, Z_n^{(s)}(\btheta)\}$, we have, for any $\varepsilon > 0$,
\begin{small}
\begin{align*}
    P\left(\sup_{\btheta \in \mathbf{\Theta}}\bigg|\frac{1}{n_1(s)}\sum_{i=1}^{n_1(s)}Z_i^{(s)}(\btheta) -\ E[Z^{(s)}(\btheta)]\bigg| > \varepsilon\right)
    &= P\left(\sup_{\btheta \in \mathbf{\Theta}}\bigg|\frac{1}{n \frac{n_1(s)}{n}}\sum_{i=1}^{n \frac{n_1(s)}{n}}Z_i^{(s)}(\btheta) -\ E[Z^{(s)}(\btheta)]\bigg| > \varepsilon\right) \\
    &= P\left(\sup_{\btheta \in \mathbf{\Theta}}\bigg|\frac{1}{n \frac{\Tilde{n}_1(s)}{n}}\sum_{i=1}^{n \frac{\Tilde{n}_1(s)}{n}}Z_i^{(s)}(\btheta) -\ E[Z^{(s)}(\btheta)]\bigg| > \varepsilon\right) \\
    &= E\left[P\left(\sup_{\btheta \in \mathbf{\Theta}}\bigg|\frac{1}{n \frac{\Tilde{n}_1(s)}{n}}\sum_{i=1}^{n \frac{\Tilde{n}_1(s)}{n}}Z_i^{(s)}(\btheta) -\ E[Z^{(s)}(\btheta)]\bigg| > \varepsilon \big| \frac{\Tilde{n}_1(s)}{n}\right)\right].
\end{align*}
\end{small}

Given the above derivation, if we can show
\begin{equation}\label{eq:lemma1}
    P\left(\sup_{\btheta \in \mathbf{\Theta}}\bigg|\frac{1}{n \frac{\Tilde{n}_1(s)}{n}}\sum_{i=1}^{n \frac{\Tilde{n}_1(s)}{n}}Z_i^{(s)}(\btheta) -\ E[Z^{(s)}(\btheta)]\bigg| > \varepsilon| \frac{\Tilde{n}_1(s)}{n}\right) \rightarrow 0 \quad \textrm{almost surely},
\end{equation}
then the dominated convergence theorem implies the desired result. To see this, since $Z^{(s)}$ is P-Glivenko-Cantelli, then $\sup_{\btheta \in \mathbf{\Theta}}\big|\frac{1}{n_k}\sum_{i=1}^{n_k} Z_i^{(s)}(\btheta) - E[Z^{(s)}(\btheta)]\big| \xrightarrow{P} 0 $ for any $n_k \rightarrow \infty$ as $k \rightarrow \infty$.
Then~(\ref{eq:lemma1}) follows from the almost sure convergence of  $n\frac{\Tilde{n}_1(s)}{n}$ to infinity and the independence of $\frac{\Tilde{n}_1(s)}{n}$ and $\{Z_1^{(s)}(\btheta), \dots, Z_n^{(s)}(\btheta)\}$.
\end{proof}
\begin{lemma}\label{lemma2}
Given Assumption 1, let $Z_i(1) = h_1(Y_i(1), M_i(1), \bX_i)$ and \\ $Z_i(0) = h_2(Y_i(0), M_i(0), \bX_i)$ for some functions $h_1$ and $h_2$ such that $E[Z_i(a)^2] < \infty$ for $a = 0, 1$. Then under stratified randomization or the biased-coin design,
\begin{displaymath}
\frac{1}{\sqrt{n}}\sum_{i=1}^n \left\{\frac{A_i}{\pi}(Z_i(1)-E[Z_i(1)]) - \frac{1-A_i}{1-\pi}(Z_i(0)-E[Z_i(0)])\right\} \xrightarrow{d} N(0, \sigma_1^2 + \sigma_2^2),
\end{displaymath}
where
\begin{align*}
    \sigma_1^2 &= \frac{1}{\pi}Var(Z(1) - E[Z(1)|S]) + \frac{1}{1-\pi}Var(Z(0) - E[Z(0)|S]),\\
    \sigma_2^2 &= Var(E[Z(1) - Z(0)|S]).
\end{align*}
\end{lemma}

\begin{proof}
See Lemma B.1 and Lemma B.2 in the supplementary material of \cite{bugni2018}. The only difference is that we replace $Y_i(a)$ by $Z_i(a)$ and all of the arguments still hold.
\end{proof}

\begin{lemma}\label{lemma:E*}
Let $f(\bX, Y(a)M(a), M(a))$ be a function with $E[f(\bX, Y(a)M(a), M(a))^2] < \infty$ for $a = 0,1$. Then
\begin{align*}
    E^*[I\{A=a\}f(\bX, YM, M)] &= \pi_a E[f(\bX, Y(a)M(a), M(a))],\\
    E^*[I\{A=a\}f(\bX, YM, M)|S] &= \pi_a E[f(\bX, Y(a)M(a), M(a))|S],
\end{align*}
where $\pi_a = \pi a + (1-\pi)(1-a)$.
\end{lemma}
\begin{proof}
We first prove the second equation.
The definition of $P^*$ implies that, for $a = 0,1$ and $s \in \mathcal{S}$, the distribution of $f(\bX, YM, M)$ given $A=a, S = s$ is equal to the distribution of $f(\bX, Y(a)M(a), M(a))$ given $A=a, S = s$. Since $A$ is independent of $(\bX, Y(a)M(a), M(a))$ and $S$ is encoded as dummy variables in $\bX$, then $A$ is independent of $S$ and $f(\bX, Y(a)M(a), M(a))$, which implies the distribution of $f(\bX, Y(a)M(a), M(a))$ given $A=a, S = s$ is the same as the distribution of $f(\bX, Y(a)M(a), M(a))$ given $S=s$. Hence $E^*[f(\bX, YM, M)|A=a, S=s] = E[f(\bX, Y(a)M(a), M(a))|S=s]$. Then the second equation is proved by the following derivation:
\begin{align*}
    &E[f(\bX, Y(a)M(a), M(a))|S=s] \\
    &=
    E^*[f(\bX, YM, M)|A=a, S=s] \\ 
    &= \frac{1}{P(A=a, S=s)}E^*[I\{A=a\}I\{S=s\}f(\bX, YM, M)] \\
    &= \frac{1}{P(A=a)P(S=s)}E^*[I\{A=a\}I\{S=s\}f(\bX, YM, M)]\\
    &= \frac{1}{P(A=a)}E^*[I\{A=a\}f(\bX, YM, M)|S=s].
\end{align*}
The first equation is followed because
\begin{align*}
    &E^*[I\{A=a\}f(\bX, YM, M)] \\
    &=\sum_{s\in \mathcal{S}}E[I\{A=a\} I\{S=s\}f(\bX, Y(a)M(a), M(a))] \\
    &= \sum_{s\in \mathcal{S}} E^*[f(\bX, YM, M)|A=a, S=s]P(A=a, S=s) \\
    &= \sum_{s\in \mathcal{S}}E[f(\bX, Y(a)M(a), M(a))|S=s]P(A=a)P(S=s) \\
    &= P(A=a)E[f(\bX, Y(a)M(a), M(a))].
\end{align*}
\end{proof}

\begin{lemma}\label{lemma:observed-data-distribution}
Given Assumption 1, under simple randomization, stratified randomization or the biased-coin design, each observed data vector $(A_i, \bX_i, Y_iM_i, M_i)$ is identically distributed and $A_i$ is independent of $\bW_i$ for $i = 1, \dots, n$. Then the observed data distribution $P^*$ is the same as first independently generating $A$ from Bernoulli$(\pi)$ and $\bW$ from distribution $P$ and then constructing $YM$ and $M$ by $Y = Y(1)A + Y(0)(1-A)$ and $M = M(1)A + M(0)(1-A)$.
\end{lemma}

\begin{proof}
Since $(A_i, \bX_i, Y_iM_i, M_i)$ is a function of $(A_i, \bW_i)$, it suffices to show $(A_i, \bW_i)$ are identically distributed and  $A_i$ is independent of $\bW_i$. For simple randomization, since $A_i$ is identically distributed and $A_i$ is independent of $\bW_i$, then Assumption 1 (i) implies the desired result.

For stratified randomization, let $b$ denote the block size for all strata. Given the definition of stratified randomization, we have
\begin{displaymath}
 P(A_i=1|\bW_i) = P(A_i=1|S=S_i) = \frac{\binom{b-1}{\pi b - 1}}{\binom{b}{\pi b}} = \pi.
\end{displaymath}
Hence $A_i$ is independent of $\bW_i$ and $A_i$ is identically distributed. Then Assumption 1 (i) implies the desired result for stratified randomization.

For the biased-coin design,  by definition, we have $\pi = 0.5$. Let $B(s,i) = \sum_{j=1}^{i}(A_j-0.5)I\{S_j=s\}$ denote the treatment imbalance in stratum $s$ for participants $1,2,\dots, i$. We have, for any $\boldsymbol{w}$ and $s\in\mathcal{S}$ such that $\bW_i = \boldsymbol{w}$ with $S_i = s$, 
\begin{align*}
 P(A_i = 1|\bW_i = \boldsymbol{w}) &= E[P(A_i = 1|A_1,\dots, A_{i-1}, \bW_1,\dots, \bW_{i-1}, \bW_i)|\bW_i = \boldsymbol{w}] \\
 &=    E[P(A_i=1|A_1,\dots,A_{i-1},S_1, \dots, S_{i-1}, S_i = s)|\bW_i = \boldsymbol{w}] \\
 &= 0.5 P\left(B(s,i-1) = 0\right)  + \lambda  P\left(B(s,i-1) < 0\right) \\
 &\quad + (1-\lambda)P\left(B(s,i-1) > 0\right),
\end{align*}
where the last equation uses the definition of the biased-coin design and the fact that $\bW_i$ is independent of $(A_1,\dots,A_{i-1},S_1, \dots, S_{i-1})$.
If we can show $P\left(B(s,i-1) > 0\right) = P\left(B(s,i-1) < 0\right)$, then direct calculation shows that $P(A_i = 1|\bW_i) = 0.5$, which implies that $A_i$ is independent of $\bW_i$ and $A_i$ is identically distributed. Then Assumption 1 (i) implies the desired result for the biased-coin design randomization. To this end, we prove a stronger result: for any $m \in \mathbb{R}$ and positive integer $i$, 
\begin{equation}\label{eq:biased-coin-induction}
    P\left(B(s,i) = m\right) = P\left(B(s,i) = -m\right).
\end{equation}
The proof is done by induction. First, equation~(\ref{eq:biased-coin-induction}) holds for $i=1$ by definition. If equation~(\ref{eq:biased-coin-induction}) holds for $i=k-1$, then
\begin{align*}
  & P\left(B(s,k)= m\bigg|S_k=s\right) \\
  &=  P\left(A_k=1\bigg|B(s,k-1) = m-0.5, S_k=s\right)P\left(B(s,k-1) = m-0.5\right) \\
  &\quad +P\left(A_k=0\bigg|B(s,k-1) = m+0.5, S_k=s\right)P\left(B(s,k-1) = m+0.5\right) \\
  &=  P\left(A_k=0\bigg|B(s,k-1) = -m+0.5, S_k=s\right)P\left(B(s,k-1) = -m+0.5\right) \\
  &\quad +P\left(A_k=1\bigg|B(s,k-1) = -m-0.5, S_k=s\right)P\left(B(s,k-1) = -m-0.5\right) \\
  &= P\left(B(s,k)= -m\bigg|S_k=s\right),
\end{align*}
where the second last equation uses the fact that $P(A_k=0|B(s,k-1)>0, S_k=s) = P(A_k=1|B(s,k-1)<0, S_k=s)$. Following a similar proof, we have $P(B(s,k)= m|S_k\ne s) = P(B(s,k)= -m|S_k\ne s)$. Hence equation~(\ref{eq:biased-coin-induction}) holds by law of total probability, which completes the proof.
\end{proof}

\begin{proof}[Proof of Theorem 1]
Under simple randomization, the results of Theorem 1 have been proved in Section 5 of \cite{vaart_1998}. Hence, it remains to prove Theorem 1 under stratified randomization and the biased-coin design,which we do below.
To simplify the notation, we define $\bpsi^{(a)}(\btheta) = \bpsi(a, \bX, Y(a), M(a); \btheta)$ and $\bpsi_i^{(a)}(\btheta) = \bpsi(a, \bX_i, Y_i(a), M_i(a); \btheta)$ for $a = 0,1$ and $i = 1,\dots, n$.

Using the fact that $Y_i = Y_i(1)A_i + Y_i(0)(1-A_i)$ and $M_i = M_i(1)A_i + M_i(0)(1-A_i)$, the estimating equations (1) from the main paper can be re-written as
\begin{displaymath}
   \frac{1}{n}\sum_{i=1}^n \bpsi(A_i,\bX_i, Y_i, M_i;\underline\btheta)= \frac{1}{n}\sum_{i=1}^n\{A_i\bpsi_i^{(1)}(\btheta) + (1-A_i)\bpsi_i^{(0)}(\btheta)\} = \mathbf{0}.
\end{displaymath}

We first show that $\widehat{\btheta} \xrightarrow{P} \underline{\btheta}$, where $\underline{\btheta}$ is the vector that solves
\begin{equation*}
\pi E[\bpsi^{(1)}(\btheta)] + (1-\pi) E[\bpsi^{(0)}(\btheta)]=\mathbf{0}.
\end{equation*}
Regularity condition (3) implies that $\underline{\btheta}$ exists and is unique. 
For each $a\in \{0,1\}$ and $s\in \mathcal{S}$, using Definition~\ref{def:conditional-process}, let $\boldsymbol{U} = (\bX^t, Y(a), M(a))^t$, $\boldsymbol{U}^{(s)}$ denote the conditional variable of $\boldsymbol{U}$ given $S=s$, and $\bpsi^{(a,s)}$ denote the conditional process of $\{\bpsi^{(a)}(\btheta): \btheta \in \boldsymbol{\Theta}\}$ given $S=s$.
Since $\boldsymbol{U}^{(s)}$ lies in the support of $\boldsymbol{U}$, then regularity condition (4) implies that the map $\btheta \mapsto \bpsi^{(a,s)}(\btheta)$ is continuous in its support and dominated by an integrable function.  
Example 19.8 of \cite{vaart_1998} implies that $\bpsi^{(a,s)}$ is P-Glivenko-Cantelli. We also have $\sup_{\btheta \in \mathbf{\Theta}}|E[\bpsi^{(a,s)}(\btheta)]| < \infty$ since $\bpsi^{(a,s)}(\btheta)$ is dominated by an integrable function.
Then Lemma 1 implies that $\sup_{\btheta \in \mathbf{\Theta}}|\frac{1}{n}\sum_{i=1}^nA_i\bpsi_i^{(a)}(\btheta) - \pi E[\bpsi^{(a)}(\btheta)]| \xrightarrow{P} 0$ and $\sup_{\btheta \in \mathbf{\Theta}}|\frac{1}{n}\sum_{i=1}^n(1-A_i)\bpsi_i^{(a)}(\btheta) - (1-\pi)E[\bpsi^{(a)}(\btheta)]| \xrightarrow{P} 0$. Hence
\begin{align*}
    & \sup_{\btheta \in \mathbf{\Theta}}\bigg|\frac{1}{n}\sum_{i=1}^n\left\{A_i\bpsi_i^{(1)}(\btheta) + (1-A_i)\bpsi_i^{(0)}(\btheta)\right\} - \left\{\pi E[\bpsi^{(1)}(\btheta)] + (1-\pi) E[\bpsi^{(0)}(\btheta)]\right\}\bigg| \\
    &\le \sup_{\btheta \in \mathbf{\Theta}}\bigg|\frac{1}{n}\sum_{i=1}^nA_i\bpsi_i^{(a)}(\btheta) - \pi E[\bpsi^{(a)}(\btheta)]\bigg| + \sup_{\btheta \in \mathbf{\Theta}}\bigg|\frac{1}{n}\sum_{i=1}^n(1-A_i)\bpsi_i^{(a)}(\btheta) - (1-\pi)E[\bpsi^{(a)}(\btheta)]\bigg| \\
    & \xrightarrow{P} 0.
\end{align*}
Since $\underline{\btheta}$ is unique and $\frac{1}{n}\sum_{i=1}^n\{A_i\bpsi_i^{(1)}(\widehat\btheta) + (1-A_i)\bpsi_i^{(0)}(\widehat\btheta)\} = \mathbf{0}$, Theorem 5.9 of \cite{vaart_1998} implies that $\widehat{\btheta} \xrightarrow{P} \btheta$.

We then show $\widehat{\btheta}$ is asymptotically linear by an argument that follows the general approach. By multivariate Taylor expansion of function $\sum_{i=1}^n \left\{	A_i\bpsi_i^{(1)}(\widehat\btheta) + (1-A_i)\bpsi_i^{(0)}(\widehat\btheta) \right\}$ around the point $\btheta = \underline{\btheta}$,  we have
\begin{align*}
    \mathbf{0} &= \frac{1}{n}\sum_{i=1}^n \left\{	A_i\bpsi_i^{(1)}(\widehat\btheta) + (1-A_i)\bpsi_i^{(0)}(\widehat\btheta) \right\}\\
    &= \frac{1}{n}\sum_{i=1}^n \left\{	A_i\bpsi_i^{(1)}(\underline\btheta) + (1-A_i)\bpsi_i^{(0)}(\underline\btheta) \right\} + \frac{1}{n}\sum_{i=1}^n \left\{A_i  \dot{\bpsi}_i^{(1)}(\underline\btheta) + (1-A_i)\dot{\bpsi}_i^{(0)}(\underline\btheta) \right\}(\widehat{\btheta} - \underline{\btheta}) \\
    &\quad + \frac{1}{2}(\widehat{\btheta} - \underline{\btheta})^t\frac{1}{n}\sum_{i=1}^n \left\{	A_i  \ddot{\bpsi}_i^{(1)}(\widetilde\btheta) + (1-A_i)\ddot{\bpsi}_i^{(0)}(\widetilde\btheta) \right\}(\widehat{\btheta} - \underline{\btheta}),
    \end{align*}
    where $\dot{\bpsi}_i^{(a)}(\underline\btheta) = \frac{\partial}{\partial\btheta}\bpsi_i^{(a)}(\btheta)\big|_{\btheta = \underline{\btheta}}$, $\ddot{\bpsi}_i^{(a)}(\widetilde\btheta) = \frac{\partial^2}{\partial\btheta\partial\btheta^t}\bpsi_i^{(a)}(\btheta)\big|_{\btheta = \widetilde{\btheta}}$
    for  $a = 0,1$  and $\widetilde{\btheta}$ is a random point on the line segment between $\widehat{\btheta}$ and $\underline{\btheta}$.
    
    According to regularity condition (5), there exists a ball $K$ around $\underline{\btheta}$ such that \\
    $\ddot{\bpsi}_i^{(a)}(\btheta)$ is dominated by a function $v(\bX_i, Y_i(a), M_i(a))$. Hence, if $\widetilde{\btheta} \in K$, then 
    \begin{align*}
    \Big|\Big|\frac{1}{n}\sum_{i=1}^n \left\{	A_i  \ddot{\bpsi}_i^{(1)}(\widetilde\btheta) + (1-A_i)\ddot{\bpsi}_i^{(0)}(\widetilde\btheta) \right\} \Big|\Big| \le \sum_{a\in \{0,1\}}\frac{1}{n}\sum_{i=1}^n ||v(\bX_i, Y_i(a), M_i(a))||,
    \end{align*}
    which is bounded in probability by the law of large numbers. Furthermore, since $P(\widetilde{\btheta} \in K) \rightarrow 1$, we then have $\frac{1}{n}\sum_{i=1}^n \left\{	A_i  \ddot{\bpsi}_i^{(1)}(\widetilde\btheta) + (1-A_i)\ddot{\bpsi}_i^{(0)}(\widetilde\btheta) \right\} = O_p(1)$.
    Recalling that $\boldsymbol{B} = E^*[\frac{\partial}{\partial \btheta} \bpsi (A, \bX, Y, M; \btheta)\big|_{\btheta = \underline{\btheta}}] = \pi E[\dot{\bpsi}^{(1)}(\underline\btheta)] + (1-\pi)E[\dot{\bpsi}^{(0)}(\underline\btheta)]$, by Lemma \ref{lemma1} and regularity condition (6), we have
    \begin{displaymath}
    \frac{1}{n}\sum_{i=1}^n \left\{A_i  \dot{\bpsi}_i^{(1)}(\underline\btheta) + (1-A_i)\dot{\bpsi}_i^{(0)}(\underline\btheta) \right\} \xrightarrow{P} \boldsymbol{B}.
    \end{displaymath}
    Combination  of above facts implies that
    \begin{equation}\label{eq:before-asymptotic-linear}
    \frac{1}{n}\sum_{i=1}^n \left\{	A_i{\bpsi}_i^{(1)}(\underline\btheta) + (1-A_i){\bpsi}_i^{(0)}(\underline\btheta) \right\} = -(\boldsymbol{B} + o_p(1))(\widehat{\btheta} - \underline{\btheta}) - (\widehat{\btheta} - \underline{\btheta})^tO_p(1)(\widehat{\btheta} - \underline{\btheta}).
    \end{equation}
    Given equation (\ref{eq:before-asymptotic-linear}), we also need $\sqrt{n}(\widehat{\btheta} - \underline{\btheta}) = O_p(1)$ to prove the asymptotic linearity.  To see this, we derive
        \begin{align*}
        &\frac{1}{\sqrt{n}}\sum_{i=1}^n \left\{	A_i{\bpsi}_i^{(1)}(\underline\btheta) + (1-A_i){\bpsi}_i^{(0)}(\underline\btheta) \right\} \\
        &= \frac{1}{\sqrt{n}}\sum_{i=1}^n \left[	A_i\{{\bpsi}_i^{(1)}(\underline\btheta) -E[{\bpsi}^{(1)}(\underline\btheta)]\} + (1-A_i)\{{\bpsi}_i^{(0)}(\underline\btheta)-E[{\bpsi}^{(0)}(\underline\btheta)]\} \right] \\
        &\qquad + \frac{1}{\sqrt{n}}\sum_{i=1}^n \{A_iE[{\bpsi}^{(1)}(\underline\btheta)] + (1-A_i)E[{\bpsi}^{(0)}(\underline\btheta)]\} \\
        &= \frac{1}{\sqrt{n}}\sum_{i=1}^n \left[	A_i\{{\bpsi}_i^{(1)}(\underline\btheta) -E[{\bpsi}^{(1)}(\underline\btheta)]\} + (1-A_i)\{{\bpsi}_i^{(0)}(\underline\btheta)-E[{\bpsi}^{(0)}(\underline\btheta)]\} \right] \\
        &\qquad + \frac{1}{\sqrt{n}}\sum_{i=1}^n (A_i - \pi)E[{\bpsi}^{(1)}(\underline\btheta) - {\bpsi}^{(0)}(\underline\btheta)] \\
        &= \frac{1}{\sqrt{n}}\sum_{i=1}^n \left[	A_i\{{\bpsi}_i^{(1)}(\underline\btheta) -E[{\bpsi}^{(1)}(\underline\btheta)]\} + (1-A_i)\{{\bpsi}_i^{(0)}(\underline\btheta)-E[{\bpsi}^{(0)}(\underline\btheta)]\} \right] + o_p(1), \numberthis \label{eq:derivation1}
    \end{align*}
    where the second last equation comes from the fact that $E[\pi{\bpsi}^{(1)}(\underline\btheta) + (1-\pi) {\bpsi}^{(0)}(\underline\btheta)] = \mathbf{0}$ and the last equation results from stratified randomization or the biased coin design, which implies $\frac{1}{\sqrt{n}}\sum_{i=1}^n (A_i - \pi) \xrightarrow{P} 0$. Letting $\bZ_i(1) = \pi{\bpsi}_i^{(1)}(\underline\btheta)$ and $\bZ_i(0) = -(1-\pi){\bpsi}_i^{(0)}(\underline\btheta)$,
    by applying Lemma \ref{lemma2}, we have $\frac{1}{\sqrt{n}}\sum_{i=1}^n \left\{	A_i{\bpsi}_i^{(1)}(\underline\btheta) + (1-A_i){\bpsi}_i^{(0)}(\underline\btheta) \right\} = O_p(1)$. Multiplying $\sqrt{n}$ on both sides of equation (\ref{eq:before-asymptotic-linear}) and using $\widehat{\btheta} - \underline{\btheta} = o_p(1)$, we get
    \begin{displaymath}
     O_p(1) = - \boldsymbol{B}\sqrt{n}(\widehat{\btheta} - \underline{\btheta}) + o_p(\sqrt{n}(\widehat{\btheta} - \underline{\btheta})).
    \end{displaymath}
    Since $\boldsymbol{B}$ is invertible (regularity condition 6), we get $\sqrt{n}(\widehat{\btheta} - \underline{\btheta}) = O_p(1)$.
    Then from equation (\ref{eq:before-asymptotic-linear}), we get
    \begin{equation}\label{eq:asymptotic linear}
    \sqrt{n}(\widehat{\btheta} - \underline{\btheta}) = - \boldsymbol{B}^{-1}\frac{1}{\sqrt{n}}\sum_{i=1}^n \bpsi(A_i,\bX_i, Y_i, M_i;\underline\btheta)+ o_p(1),
    \end{equation}
    this completes the proof of asymptotic linearity.
    
    We next show $\widehat{\btheta}$ is asymptotically normal using Lemma~\ref{lemma2}. Combining expressions (\ref{eq:derivation1}) and (\ref{eq:asymptotic linear}), we have
    \begin{displaymath}
     \sqrt{n}(\widehat{\btheta} - \underline{\btheta}) = - \boldsymbol{B}^{-1}\frac{1}{\sqrt{n}}\sum_{i=1}^n \left[	A_i\{{\bpsi}_i^{(1)}(\underline\btheta) -E[{\bpsi}^{(1)}(\underline\btheta)]\} + (1-A_i)\{{\bpsi}_i^{(0)}(\underline\btheta)-E[{\bpsi}^{(0)}(\underline\btheta)]\} \right] + o_p(1).
    \end{displaymath}
    Letting $\bZ_i(1), \bZ_i(0)$ be the first entry of $-\pi\boldsymbol{B}^{-1}{\bpsi}_i^{(1)}(\underline\btheta)$ and $ (1-\pi)\boldsymbol{B}^{-1}{\bpsi}_i^{(0)}(\underline\btheta)$ respectively,
    we apply Lemma \ref{lemma2} to the first entry of $\sqrt{n}(\widehat{\btheta}-\underline{\btheta})$ and get $\sqrt{n}(\widehat{\Delta} - \underline{\Delta}) \xrightarrow{d} N(0, V^*)$, where $V^*$ is the first-row, first-column entry of $\boldsymbol{B}^{-1}\boldsymbol{D}\boldsymbol{B}^{-1,t}$ with
    \begin{align*}
        \boldsymbol{D} &= \pi E[Var\{{\bpsi}^{(1)}(\underline\btheta)|S\}] + (1-\pi)E[Var\{{\bpsi}^{(0)}(\underline\btheta)|S\}] \\
        &\qquad + E\left[E[\pi{\bpsi}^{(1)}(\underline\btheta) + (1-\pi){\bpsi}^{(0)}(\underline\btheta)|S]E[\pi{\bpsi}^{(1)}(\underline\btheta) + (1-\pi){\bpsi}^{(0)}(\underline\btheta)|S]^t\right],
    \end{align*}
    where $Var$ represents the variance-covariate matrix.
    
    To show that $V = V^*$, we use the result by Theorem 5.21 of \cite{vaart_1998} that $\widetilde{V}$ is the first-row first-column entry of $\boldsymbol{B}^{-1}\boldsymbol{C}\boldsymbol{B}^{-1,t}$, where, by Lemma~\ref{lemma:E*},
    \begin{align*}
    \boldsymbol{C} &= E^*[\bpsi(A,\bX, Y, M;\underline\btheta)\bpsi(A,\bX, Y, M;\underline\btheta)^t] \\
    &=   \pi E[{\bpsi}^{(1)}(\underline\btheta){\bpsi}^{(1)}(\underline\btheta)^t] + (1-\pi)E[{\bpsi}^{(0)}(\underline\btheta){\bpsi}^{(0)}(\underline\btheta)^t].
    \end{align*}
    Then it suffices to show that the first-row first-column entry of $\boldsymbol{B}^{-1}(\boldsymbol{C} - \boldsymbol{D})\boldsymbol{B}^{-1,t}$ is $\frac{1}{\pi(1-\pi)}E^*\left[E^*[(A-\pi)IF(A,\bX,Y,M)|S]^2\right]$, where $IF(A,\bX,Y,M)$ is the first entry of \\ $-\boldsymbol{B}^{-1}\bpsi(A,\bX, Y, M;\underline\btheta)$. We have the following derivation:
    \begin{align*}
      \boldsymbol{C} - \boldsymbol{D}
      &=  \pi E[{\bpsi}^{(1)}(\underline\btheta){\bpsi}^{(1)}(\underline\btheta)^t] + (1-\pi)E[{\bpsi}^{(0)}(\underline\btheta){\bpsi}^{(0)}(\underline\btheta)^t]\\
      &\quad - \pi \left(E[{\bpsi}^{(1)}(\underline\btheta){\bpsi}^{(1)}(\underline\btheta)^t] - E\left[E[{\bpsi}^{(1)}(\underline\btheta)|S]E[{\bpsi}^{(1)}(\underline\btheta)|S]^t\right]\right) \\
      &\qquad - (1-\pi)\left(E[{\bpsi}^{(0)}(\underline\btheta){\bpsi}^{(0)}(\underline\btheta)^t] - E\left[E[{\bpsi}^{(0)}(\underline\btheta)|S]E[{\bpsi}^{(0)}(\underline\btheta)|S]^t\right]\right)\\
      &\qquad - E\left[E[\pi{\bpsi}^{(1)}(\underline\btheta) + (1-\pi){\bpsi}^{(0)}(\underline\btheta)|S]E[\pi{\bpsi}^{(1)}(\underline\btheta) + (1-\pi){\bpsi}^{(0)}(\underline\btheta)|S]^t\right] \\
      &= \pi E\left[E[{\bpsi}^{(1)}(\underline\btheta)|S]E[{\bpsi}^{(1)}(\underline\btheta)|S]^t\right] + (1-\pi) E\left[E[{\bpsi}^{(0)}(\underline\btheta)|S]E[{\bpsi}^{(0)}(\underline\btheta)|S]^t\right]\\
      &\qquad -E\left[E[\pi{\bpsi}^{(1)}(\underline\btheta) + (1-\pi){\bpsi}^{(0)}(\underline\btheta)|S]E[\pi{\bpsi}^{(1)}(\underline\btheta) + (1-\pi){\bpsi}^{(0)}(\underline\btheta)|S]^t\right]\\
      &= \pi(1-\pi)E\left[E[{\bpsi}^{(1)}(\underline\btheta) - {\bpsi}^{(0)}(\underline\btheta)|S]E[{\bpsi}^{(1)}(\underline\btheta) - {\bpsi}^{(0)}(\underline\btheta)|S]^t\right] \\
      &= \frac{1}{\pi(1-\pi)}E^*\left[E^*[(A-\pi)\bpsi(A,\bX, Y, M;\underline\btheta)|S]E^*[(A-\pi)\bpsi(A,\bX, Y, M;\underline\btheta)|S]^t\right].
    \end{align*}
    The last equation uses Lemma~\ref{lemma:E*}, which implies that $E^*[(A-\pi)\bpsi(A,\bX, Y, M;\underline\btheta)|S] = E[{\bpsi}^{(1)}(\underline\btheta) - {\bpsi}^{(0)}(\underline\btheta)|S]$ and $E[f(S)] = E^*[f(S)]$ for any $f(S)$ with finite second moment. Then using the definition of $IF(A,\bX,Y,M)$, the first-row first-column entry of $\boldsymbol{B}^{-1}(\boldsymbol{C} - \boldsymbol{D})\boldsymbol{B}^{-1,t}$ is $\frac{1}{\pi(1-\pi)}E^*\left[E^*[(A-\pi)IF(A,\bX,Y,M)|S]^2\right]$. This completes the proof of asymptotic normality with the desired asymptotic variance.
    
    To prove that $V$ can be consistently estimated by $\widehat{V}$ defined in Section \ref{app:variance-estimator}, we first show $\widehat{\boldsymbol{B}} \xrightarrow{P} \boldsymbol{B}$.
    By multivariate Taylor's expansion of $\frac{1}{n\pi} \sum_{i=1}^nA_i\dot{\bpsi}_i^{(1)}(\widehat\btheta)$ around the point $\btheta = \underline{\btheta}$, we have
    \begin{align*}
        & \frac{1}{n\pi} \sum_{i=1}^nA_i\dot{\bpsi}_i^{(1)}(\widehat\btheta) - \frac{1}{n\pi}\sum_{i=1}^nA_i\dot{\bpsi}_i^{(1)}(\underline\btheta)
        = \frac{1}{n\pi}\sum_{i=1}^nA_i\ddot{\bpsi}_i^{(1)}(\widetilde\btheta)(\widehat{\btheta} - \underline{\btheta}),
    \end{align*}
    where $\widetilde{\btheta}$ is a random point on the line segment between $\widehat{\btheta}$ and $\underline{\btheta}$.
Regularity condition (5) indicates that $\frac{1}{n}\sum_{i=1}^nA_i\ddot{\bpsi}_i^{(1)}(\widetilde\btheta) = O_p(1)$ since $\widehat{\btheta} \xrightarrow{P} \underline{\btheta}$. As a result, by Lemma~\ref{lemma1},
\begin{align*}
    \frac{1}{n\pi}\sum_{i=1}^nA_i\dot{\bpsi}_i^{(1)}(\widehat\btheta)
    = \frac{1}{n\pi}\sum_{i=1}^nA_i\dot{\bpsi}_i^{(1)}(\underline\btheta) + o_p(1) = E\left[\dot{\bpsi}^{(1)}(\underline\btheta)\right] + o_p(1).
\end{align*}
Similarly, we have
$ \sum_{i=1}^n\frac{1-A_i}{n(1-\pi)}\dot{\bpsi}_i^{(0)}(\widehat\btheta) = E\left[\dot{\bpsi}^{(0)}(\underline\btheta)\right] + o_p(1).$
Hence
\begin{align*}
    \widehat{\boldsymbol{B}} &= \frac{1}{n}\sum_{i=1}^n\frac{\partial}{\partial\btheta} \bpsi(A_i, \bX_i, Y_i, M_i; \btheta) \bigg|_{\btheta = \widehat{\btheta}} \\
    &= \frac{1}{n}\sum_{i=1}^nA_i\dot{\bpsi}_i^{(1)}(\widehat\btheta) + \frac{1}{n}\sum_{i=1}^n(1-A_i)\dot{\bpsi}_i^{(0)}(\widehat\btheta) \\
    &= \pi E\left[\dot{\bpsi}^{(1)}(\underline\btheta)\right] + (1-\pi)E\left[\dot{\bpsi}^{(0)}(\underline\btheta)\right] + o_p(1) \\
    &= \boldsymbol{B} + o_p(1).
\end{align*}
Following a similar proof, Lemma \ref{lemma1} implies $\widehat{\boldsymbol{C}} \xrightarrow{P} \boldsymbol{C}$ and $\widehat{d}(s) \xrightarrow{P} E^*[(A-\pi) \bpsi(A_i, \bX_i, Y_i, M_i)|S = s]$ for each $s \in \mathcal{S}$. By continuous mapping theorem, we get $\widehat{V} \xrightarrow{P} V$.
\end{proof}

\subsection{Proof of Corollary 1}\label{app:proofs-corollary1}

\begin{proof}
Since the ANCOVA estimator and standardized logistic regression estimator are special cases of the DR-WLS estimator, we only present the proof for the DR-WLS estimator.
Given Assumption 1 and regularity conditions in Section~\ref{app:regularity-conditions}, Theorem 1 shows that  $\widehat{\Delta}_{DR-WLS}$ is consistent to $\underline{\Delta}$ and asymptotically normal.

We first show that $\widehat{\Delta}_{DR-WLS}$ is doubly robust. Denote $h(A,\bX) = g^{-1}(\underline\beta_0 + \underline\beta_AA + \underline\bbeta_{\bX}^t \bX)$ and $e(A,\bX) = \mbox{expit}(\underline\alpha_0 + \underline\alpha_AA + \underline\balpha_{\bX}^t\bX)$. If we assume that the model for missingness is correctly specified, then $E[M(a)|\bX] = e(a,\bX)$ for $a = 0, 1$. Using the missing at random assumption and the positivity assumption, we have $$\resizebox{\textwidth}{!}{$0 = E[M(a)\frac{Y(a) - h(a,\bX)}{e(a,\bX)}] = E\left[E[M(a)|\bX]\frac{E[Y(a)|\bX] - h(a,\bX)}{e(a,\bX)}\right] = E[Y(a)] - E[h(a,\bX)]$.}$$ 
This implies that $E[Y(a)]=E[h(a,\bX)]$ for each $a=0,1$; this claim also holds if, instead of assuming that the model for missingness is correctly specified, we assume that $h(A, \bX)$ is correctly specified, then $E[Y(a)|\bX] = h(a,\bX)$ and hence $E[Y(a)] = E[E[Y(a)|\bX]] =  E[h(a,\bX)]$ for $a = 0, 1$. As a result, if at least one of the two working models is correctly specified, then $\underline{\Delta} = E[h(1,\bX)] - E[h(0,\bX)] = \Delta^*$. This proves double robustness of the DR-WLS estimator.

We next calculate $\widetilde{V} - V$. Derived from Theorem 5.21 of \cite{vaart_1998}, the influence function of the DR-WLS estimator is
\begin{align*}
    IF(A,\bX,Y,M) &= \left(\frac{A-\pi}{\pi(1-\pi)} - \boldsymbol{c}_1 \bZ\right)\frac{M\{Y-h(A,\bX)\}}{ e(A,\bX)} - \boldsymbol{c}_2 \bZ \{M - e(A,\bX)\} \\
    &\qquad + h(1,\bX) - h(0,\bX) - \underline\Delta,
\end{align*}
where $\bZ = (1, A, \bX^t)^t$,
\begin{align*}
    \boldsymbol{c}_1  &= E^*\left[\frac{A-\pi}{\pi(1-\pi)}\left(\frac{M}{e(A,\bX)}-1\right)h_{\underline{\beta}}(A,\bX) \right]^t \left\{E^*\left[\frac{M}{e(A,\bX)}h_{\underline{\beta}}(A,\bX)\bZ^t\right]\right\}^{-1},\\
    \boldsymbol{c}_2 &=  E[h_{\underline{\beta}}(1,\bX) - h_{\underline{\beta}}(0,\bX)]^t\left\{E^*\left[\frac{M}{e(A,\bX)}h_{\underline{\beta}}(A,\bX)\bZ^t\right]\right\}^{-1} \\
    &\qquad E^*\left[\frac{Me_{\underline{\alpha}}(A,\bX)}{e(A,\bX)^2}\{Y - h(A,\bX)\}\bZ^t\right]\{E^*[ e_{\underline{\alpha}}(A,\bX)\bZ^t]\}^{-1},
\end{align*}
with $h_{\underline{\beta}}(A,\bX) = \frac{\partial}{\partial \beta}  g(\beta_0  + \beta_AA + \bbeta_{\bX}^t\bX)\big|_{\beta = \underline{\beta}},e_{\underline{\alpha}}(A,\bX) = \frac{\partial}{\partial \alpha} \mbox{expit}(\alpha_0 + \alpha_AA + \balpha_{\bX}^t\bX)\big|_{\alpha = \underline{\alpha}}$.

Applying Theorem 1, we get $\widetilde{V} - V = \frac{1}{\pi(1-\pi)}E^*\left[E^*[(A-\pi)IF(A,\bX,Y,M)|S]^2\right]$. To calculate this quantity, we observe that, if the outcome regression model is correctly specified, then $\boldsymbol{c}_2  = \mathbf{0}$  and $E^*[(A-\pi) M\frac{Y - h(A,\bX)}{e(A,\bX)}\bZ|S] = \mathbf{0}$; if the missing model is correctly specified, then $\boldsymbol{c}_1  = \mathbf{0}$  and $E^*[(A-\pi)\{M - e(A, \bX)\}\bZ|S] = \mathbf{0}$. Furthermore, Lemma~\ref{lemma:E*} implies that $E^*[(A-\pi)\{h(1,\bX) - h(0,\bX) - \underline\Delta\}|S] = 0$.
Hence, if at least one of the two models is correct, then 
\begin{displaymath}
 E^*[(A-\pi)IF(A,\bX,Y,M)|S] = E^*\left[\frac{(A-\pi)^2}{\pi(1-\pi)}\frac{M\{Y-h(A,\bX)\}}{ e(A,\bX)}\bigg | S\right].
\end{displaymath}
Since $S$ is a categorical variable and adjusted in the outcome regression model, then the definition of $\underline{\btheta}$ (i.e., the solution to $E^*[\bpsi(A,\bX, Y, M;\btheta)] = \mathbf{0}$) implies that $E^*\left[\frac{M\{Y-h(A,\bX)\}}{ e(A,\bX)} \big| S\right] = 0$. Therefore
\begin{displaymath}
 \widetilde{V} - V = \frac{(1-2\pi)^2}{\pi^3(1-\pi)^3} E^*\left[E^*\left[\frac{AM\{Y-h(A,\bX)\}}{ e(A,\bX)}\bigg|S\right]^2\right].
\end{displaymath}
When $\pi = 0.5$, it is straightforward that $\widetilde{V} = V$. If the outcome regression model includes treatment-by-randomization-strata interaction terms, then according to the definition of $\underline{\btheta}$,  $E^*\left[\frac{AM\{Y-h(A,\bX)\}}{ e(A,\bX)}I\{S = s\}\right] = 0$ for each $s \in \mathcal{S}$, we have $E^*\left[\frac{AM\{Y-h(A,\bX)\}}{ e(A,\bX)} \big| S\right] = 0$ and hence $\widetilde{V} = V$. If the outcome regression model is correctly specified, then $E[Y(1)|\bX] = h(1,\bX)$. Together with the missing at random assumption, we have $E\left[\frac{M(1)\{Y(1)-h(1,\bX)\}}{ e(1,\bX)} \big| \bX\right] = E\left[\frac{E[M(1)|\bX]\{E[Y(1)|\bX]-h(1,\bX)\}}{ e(1,\bX)} \big| \bX\right] = 0$. By Lemma~\ref{lemma:E*}, we have  $E^*\left[\frac{AM\{Y-h(A,\bX)\}}{ e(A,\bX)} \big| S\right] = \pi E\left[\frac{M(1)\{Y(1)-h(1,\bX)\}}{ e(1,\bX)} \big| S\right] = 0$, which implies $\widetilde{V} = V$.
\end{proof}

\subsection{Proof of Theorem 2}\label{app:proofs-thm2}

In this section, we first present Lemma 4, which gives a central limit theorem for dependent stochastic processes under stratified randomization or the biased-coin design. We then present Lemma 5, which gives detailed algebra for a decomposition used in the proof of Lemma 4. Given Lemma 4, we prove Theorem 2. 

\begin{lemma}\label{lemma1-KM}
Let $S$ be a random variable taking values in a discrete set $\mathcal{S} = \{1, \dots, K\}$. Let $ R = \{R(t): t \in [0,\tau]\}$ be a real-valued, uniformly bounded (i.e. $P(\sup_t |R(t)| < M) = 1$ for some constant $M>0$) stochastic process and $R^{(s)}$ be the conditional process of $R$ given $S=s$ (defined by Definition~\ref{def:conditional-process}). We assume $R^{(s)}$ is P-Donsker for each $s \in \mathcal{S}$. Let $(S_i, R_i), i = 1,\dots, n$ be independent, identically distributed samples from the joint distribution of $(S, R)$.
Let $(A_1, \dots, A_n)$ be a vector of binary random variables such that (1) $(R_1, \dots, R_n) \indep (A_1, \dots, A_n)|(S_1, \dots, S_n)$ and (2) $n^{-1/2}\sum_{i=1}^n(A_i - \pi)I\{S_i = s\}|(S_1, \dots, S_n) \xrightarrow{P} 0$ a.s. for every $s \in \mathcal{S}$ and some constant $\pi \in (0,1)$.
Then $\frac{1}{\sqrt{n}}\sum_{i=1}^n\left(\frac{A_i}{\pi}R_i(t) - E[R(t)]\right)$ weakly converges to a tight, mean $0$ Gaussian process with covariance function
\begin{displaymath}
V^{(a)}(t, t') = \frac{1}{\pi}E[Cov(R(t), R(t')|S)] + Cov(E[R(t)|S], E[R(t')|S]).
\end{displaymath}
\end{lemma}
\begin{proof}
Consider the following derivation, which uses the same technique as Lemma B.1 of the supplementary material of \cite{bugni2018} with generalization to stochastic processes:
\begin{displaymath}
\frac{1}{\sqrt{n}}\sum_{i=1}^n\left(\frac{A_i}{\pi}R_i(t) - E[R(t)]\right) = U_{n,1}(t) + U_{n,2}(t) + U_{n,3}(t),
\end{displaymath}
where
\begin{align*}
     U_{n,1}(t)&= \frac{1}{\sqrt{n}}\sum_{i=1}^n\frac{A_i}{\pi}\left(R_i(t) - E[R(t)|S=S_i]\right), \\
     U_{n,2}(t) &=\sum_{s\in\mathcal{S}} \sqrt{n}(\frac{n(s)}{n} - p(s))(E[R(t)|S = s] - E[R(t)]), \\
     U_{n,3}(t)&= \sum_{s\in\mathcal{S}} \frac{D_n(s)}{\pi \sqrt{n}}E[R(t)|S = s],
\end{align*}
with $D_n(s) = \sum_{i=1}^n(A_i - \pi)I\{S_i = s\}$, $n(s) = \sum_{i=1}^nI\{S_i = s\}$ and $p(s) = P(S=s)$. We prove the above derivation in Lemma~\ref{lemma2-km}. In the rest of the proof, we use $U_{n,1}, U_{n,2}, U_{n,3}$ to represent the processes $\{U_{n,1}(t):t\in[0,\tau]\}$, $\{U_{n,2}(t):t\in[0,\tau]\}$ and $\{U_{n,3}(t):t\in[0,\tau]\}$ respectively.

We first show that $U_{n,3}(t) \xrightarrow{d} 0$ uniformly in $t$. 
Since the definition of $(A_1, \dots, A_n)$
implies that $\frac{D_n(s)}{\sqrt{n}} \xrightarrow{d} 0$, then for any $\varepsilon > 0$, 
\begin{align*}
    P^*\left(\sup_t|U_{n,3}(t)| > \varepsilon\right) \le P^*\left(\sum_{s\in\mathcal{S}}\sup_t\bigg|E[R(t)|S = s]\bigg|\bigg|\frac{D_n(s)}{\pi \sqrt{n}}\bigg| > \varepsilon\right),
\end{align*}
where $P^*$ represents outer probability measure (Section 6.2 of \citealp{kosorok2008}).
Since $R(t)$ is uniformly bounded, we can bound $\sup_t\big|E[R(t)|S = s]\big|$ by a constant for all $s \in \mathcal{S}$. Then we get $\lim_{n\rightarrow \infty} P^*\left(\sup_t|U_{n,3}(t)| > \varepsilon\right) = 0$.


We next show that the process $U_{n,2}$ weakly converges to a tight, mean 0 Gaussian process with covariance function $V_2(t,t') = Cov(E[R(t)|S], E[R(t')|S])$. For each $s \in \mathcal{S}$, since $\{I\{S = s\}:t\in [0,\tau]\}$ is P-Donsker and $E[R(t)|S = s] - E[R(t)]$ is uniformly bounded, then $\{\sum_{s\in\mathcal{S}}I\{S = s\}(E[R(t)|S = s] - E[R(t)]):t\in [0,\tau]\}$ is P-Donsker by Corollary 9.32 (i) and (v) of \cite{kosorok2008}. 
Hence $U_{n,2}$ weakly converges to a tight, mean 0 Gaussian process. The covariance function can be derived accordingly.

We next construct a process $U^*_{n,1}$ such that
$U^*_{n,1}$ is independent of $(U_{n,2}, U_{n,3})$ and $U^*_{n,1}$ weakly converges to a tight, mean 0 Gaussian process with covariance function
\begin{displaymath}
 V_1(t,t') = \frac{1}{\pi}E[Cov(R(t), R(t')|S)].
\end{displaymath}
We then show in the next paragraph that $U_{n,1}(t) \overset{d}{=} U^*_{n,1}(t) + o_{p^*}(1)$, where $\overset{d}{=}$ means random variables on both sides have the same distribution and  $o_{p^*}(1)$ represents a sequence of processes $\{X(t):t\in[0,\tau]\}$ such that $\lim_{n\rightarrow\infty}P^*(\sup_{t\in[0,\tau]}|X_n(t)|>\varepsilon) = 0$ for any $\varepsilon>0$.
To this end, independently for each $s \in \mathcal{S}$ and independently of $(A_1, \dots, A_n)$ and $(S_1, \dots, S_n)$, let $\{R_i^{(s)}(t): t\in[0,\tau]\},i = 1\dots, n$ be independent and identically distributed samples from process $R^{(s)}$.
We define
\begin{displaymath}
U^*_{n,1}(t) = \sum_{s \in \mathcal{S}}\left(\frac{1}{\pi\sqrt{n}}\sum^{\lfloor n(F(s)+ \pi p(s))\rfloor}_{i=\lfloor nF(s)\rfloor+1}\{R_i^{(s)}(t) - E[R_i^{(s)}(t)]\}\right),
\end{displaymath}
where $F(s) = P(S_i \le s)$. 
Since $U_{n,1}^*(t)$ is a function of $R_i^{(s)}(t)$, then $U_{n,1}^*(t)$ is independent of $(A_1,S_1, \dots, A_n,S_n)$. Using the fact that $U_{n,2}(t), U_{n,3}(t)$ are functions of $(A_1,S_1, \dots, A_n,S_n)$, we conclude that $U_{n,1}^*$ is independent of $(U_{n,2}, U_{n,3})$. 
For $x \in \mathbb{R}$, let $\lfloor x \rfloor$ be the largest integer smaller than or equal to $x$.
Since $R^{(s)}$ is P-Donsker and $R_i^{(s)}(t), i \in \lfloor nF(s)\rfloor+1, \dots, \lfloor n(F(s)+ \pi p(s))\rfloor$ are i.i.d with $\lfloor n\pi p(s)\rfloor$ (or $\lfloor n\pi p(s)\rfloor + 1$) units for each $s \in \mathcal{S}$, we have $\frac{1}{\pi\sqrt{n}}\sum^{\lfloor n(F(s)+ \pi p(s))\rfloor}_{i=\lfloor nF(s)\rfloor+1}\{R_i^{(s)}(t) - E[R_i^{(s)}(t)]\}$ converges weakly to a tight Gaussian process with covariance function $\frac{p(s)}{\pi}Cov(R(t),R(t')|S=s)$. Since the data from different strata are independent, we get the desired convergence of $U^*_{n,1}(t)$.


We next show $U_{n,1}(t) \overset{d}{=} U^*_{n,1}(t) + o_{p^*}(1)$.
Given the definition of $(A_1, \dots, A_n)$, $U_{n,1}(t)$ has the same distribution of the same quantity where data are ordered by strata and then by treatment group ($A_i=1$ first) within each stratum. 
Define
\begin{displaymath}
\widetilde{U}_{n,1}(t) = \sum_{s \in \mathcal{S}}\left(\frac{1}{\pi\sqrt{n}}\sum^{N(s) + n_1(s)}_{i=N(s)+1}\{R_i^{(s)}(t) - E[R_i^{(s)}(t)]\} \right),
\end{displaymath}
where $N(s) = \sum_{i=1}^nI\{S_i \le s\}$ and $n_1(s) = \sum_{i=1}^nI\{S_i \le s, A_i = 1\}$. 
By the definition of $R_i^{(s)}(t)$, we have $\{U_{n,1}(t) |A_1,S_1, \dots, A_n,S_n\}\overset{d}{=}\{\widetilde{U}_{n,1}(t) |A_1,S_1, \dots, A_n,S_n\}$ and hence $U_{n,1}(t)\overset{d}{=}\widetilde{U}_{n,1}(t)$. 
It suffices to prove that $\widetilde{U}_{n,1}(t) = U^*_{n,1}(t) + o_{p^*}(1)$, that is, for each fixed $s \in \mathcal{S}$,
\begin{displaymath}
 \Delta_{n,s}(t) = \frac{1}{\sqrt{n}}\left(\sum^{N(s) + n_1(s)}_{i=N(s)+1} - \sum^{\lfloor n(F(s)+ \pi p(s))\rfloor}_{i=\lfloor nF(s)\rfloor+1}\right)\{R_i^{(s)}(t) - E[R_i^{(s)}(t)]\}
\end{displaymath}
converges in distribution to $0$ uniformly. 
To this end, we observe that $|\Delta_{n,s}(t)| \le \widetilde\Delta_{n,s}(t) $, where 
\begin{align*}
     \widetilde\Delta_{n,s}(t) &= \frac{1}{\sqrt{n}}\bigg|\sum_{\min\{N(s), \lfloor nF(s)\rfloor\} + 1}^{\max\{N(s), \lfloor nF(s)\rfloor\}}\{R_i^{(s)}(t) - E[R_i^{(s)}(t)]\}\bigg|\\
     &\quad + \frac{1}{\sqrt{n}}\bigg|\sum_{\min\{N(s)+n_1(s), \lfloor nF(s)+n\pi p(s)\rfloor\} + 1}^{\max\{N(s)+n_1(s), \lfloor nF(s)+n\pi p(s)\rfloor\}}\{R_i^{(s)}(t) - E[R_i^{(s)}(t)]\}\bigg|.
\end{align*}
Define $M(s) = |N(s) - \lfloor nF(s)\rfloor|$, and $G_n^{(s)} = \{\frac{1}{\sqrt{n}}\sum_{i=1}^{n}\{\widetilde{R}_i^{(s)}(t) - E[\widetilde{R}_i^{(s)}(t)]\}:t\in [0,\tau]\}$, where $(\widetilde{R}^{(s)}_1, \dots, \widetilde{R}^{(s)}_n)$ are independent, identically distributed samples from process $R^{(s)}$ and are independent of $(R^{(s)}_1, \dots, R^{(s)}_n)$. We also define $L(s) = |N(s) + n_1(s) - \lfloor n(F(s)+ \pi p(s))\rfloor|$ and $\widetilde{G}^{(s)}_n$ to be identically distributed as $G_n^{(s)}$ and independent of $G_n^{(s)}$ and $(R^{(s)}_1, \dots, R^{(s)}_n)$. Then $\widetilde\Delta_{n,s}(t) \overset{d}{=} \sqrt{\frac{M(s)}{n}} |G^{(s)}_{M(s)}(t)| + \sqrt{\frac{L(s)}{n}} |\widetilde{G}^{(s)}_{L(s)}(t)|$. For any $\varepsilon >0$, since
\begin{align*}
  P^*(\sup_{t}|\Delta_{n,s}(t)| > \varepsilon) &\le P^*(\sup_{t}\widetilde\Delta_{n,s}(t) > \varepsilon) \\
  &\le P^*\left(\sqrt{\frac{M(s)}{n}} |G^{(s)}_{M(s)}(t)| > \frac{\varepsilon}{2}\right) + P^*\left(\sqrt{\frac{L(s)}{n}} |G^{(s)}_{L(s)}(t)| > \frac{\varepsilon}{2}\right),   
\end{align*}
it suffices to show $\sqrt{\frac{M(s)}{n}}G^{(s)}_{M(s)}(t) = o_{p^*}(1)$ and $\sqrt{\frac{L(s)}{n}} G^{(s)}_{L(s)}(t)= o_{p^*}(1)$.
The central limit theorem implies $(N(s) - nF(s))/\sqrt{n} \xrightarrow{d} N(0, F(s)(1-F(s))$. Hence for any $\delta > 0$, there exist $C_0, C_1 >0$ such that $P(\sqrt{n}C_0 \le M(s) \le \sqrt{n}C_1) \ge 1 - \delta$ for $n$ large enough. 
Since $R^{(s)}$ is P-Donsker, then $G_n^{(s)}$ weakly converges to a tight Gaussian process $G^{(s)}$. By the continuous mapping theorem (Theorem 7.7 of \citealp{kosorok2008}), we have $\sup_t|G_n(t)|$ converges weakly to $\sup_t|G^{(s)}(t)|$, which implies that, for any $\delta >0$, there exists a $C > 0$ such that $ P^*\left(\sup_t|G_n^{(s)}(t)|> C\right) < \delta $  for $n$ large enough.
Since $M(s)$ is independent of $\widetilde{R}_i^{(s)}$ and $M(s)$ takes values on a finite set for fixed $n$, by Fubini's Theorem (Lemma 1.2.7 of \citealp{vaart1996}) and the definition of iterative outer expectations (Page 10-11 of \citealp{vaart1996}), we have for any $m \ge 0$ that $P^*(\sup_t|\sqrt{\frac{m}{n}}G^{(s)}_m(t)|> \varepsilon, M(s) = m) = P^*(\sup_t|\sqrt{\frac{m}{n}}G^{(s)}_m(t)|> \varepsilon)P(M(s) = m)$. Then for any $\varepsilon, \delta>0$ and $n$ large enough,  we have the following derivation:
\begin{align*}
 & P^*\left(\sup_t|\sqrt{\frac{M(s)}{n}}G^{(s)}_{M(s)}(t)|> \varepsilon\right) \\
&\le P^*\left(\sup_t|\sqrt{\frac{M(s)}{n}}G^{(s)}_{M(s)}(t)|> \varepsilon, M(s) \in [\sqrt{n}C_0, \sqrt{n}C_1]\right) \\
&\quad\quad + P^*\left(\sup_t|\sqrt{\frac{M(s)}{n}}G^{(s)}_{M(s)}(t)|> \varepsilon,M(s) \notin [\sqrt{n}C_0, \sqrt{n}C_1]\right)
\\ 
&\le \sum_{m \in [\sqrt{n}C_0, \sqrt{n}C_1]}P^*\left(\sup_t|\sqrt{\frac{m}{n}}G^{(s)}_m(t)|> \varepsilon\right)P(M(s) = m) + P(M(s) \notin [\sqrt{n}C_0, \sqrt{n}C_1]) \\
&< \sum_{m \in [\sqrt{n}C_0, \sqrt{n}C_1]}P^*\left(\sup_t|G^{(s)}_m(t)|> n^{\frac{1}{4}}\frac{\varepsilon}{C_1}\right)P(M(s) = m) + \delta \\
&< \sum_{m \in [\sqrt{n}C_0, \sqrt{n}C_1]}\delta P(M(s) = m) + \delta \\
&\le 2 \delta.
\end{align*}
Using the same technique, we have, for $n$ large enough, $P^*(\sup_t|\sqrt{\frac{L(s)}{n}}\widetilde{G}_{L(s)}|> \varepsilon) \le 2\delta$,
which completes the proof of $U_{n,1}(t) \overset{d}{=} U^*_{n,1}(t) + o_{p^*}(1)$.

We next show $(U_{n,1}, U_{n,2}, U_{n,3}) \overset{d}{=} (U^*_{n,1}, U_{n,2}, U_{n,3}) + o_{p^*}(1)$. Since we have shown that $\{U_{n,1}(t) |A_1,S_1, \dots, A_n,S_n\}\overset{d}{=}\{\widetilde{U}_{n,1}(t) |A_1,S_1, \dots, A_n,S_n\}$ and $U_{n,2}(t), U_{n,3}(t)$ are functions of $\{A_1,S_1, \dots, A_n,S_n\}$, then $(U_{n,1}, U_{n,2}, U_{n,3}) \overset{d}{=} (\widetilde{U}_{n,1}, U_{n,2}, U_{n,3})$. Then it suffices to show that $(\widetilde{U}_{n,1}, U_{n,2}, U_{n,3}) \overset{d}{=} (U^*_{n,1}, U_{n,2}, U_{n,3}) + o_{p^*}(1)$. This is implied by the following derivation:
\begin{displaymath}
 P^*(\sup_t||(\widetilde{U}_{n,1}, U_{n,2}, U_{n,3}) - (U^*_{n,1}, U_{n,2}, U_{n,3})|| > \varepsilon) =  P^*(\sup_t|\widetilde{U}_{n,1} - U^*_{n,1}| > \varepsilon) \rightarrow 0,
\end{displaymath}
where $||\cdot||$ represents the Euclidean norm.

Finally, using the fact that $U^*_{n,1} \indep U_{n,2}$, we have
\begin{align*}
    \frac{1}{\sqrt{n}}\sum_{i=1}^n\left(\frac{A_i}{\pi}R_i(t) - E[R(t)]\right) &= U_{n,1}(t) + U_{n,2}(t) + U_{n,3}(t) \\
    &\overset{d}{=} U^*_{n,1}(t) + U_{n,2}(t) + o_{p^*}(1)
\end{align*}
weakly converges to the desired Gaussian process.
\end{proof}
\begin{lemma}\label{lemma2-km}
Consider the same setting as Lemma~\ref{lemma1-KM}. Then 
\begin{displaymath}
\frac{1}{\sqrt{n}}\sum_{i=1}^n\left(\frac{A_i}{\pi}R_i(t) - E[R(t)]\right) = U_{n,1}(t) + U_{n,2}(t) + U_{n,3}(t),
\end{displaymath}
where
\begin{align*}
     U_{n,1}(t)&= \frac{1}{\sqrt{n}}\sum_{i=1}^n\frac{A_i}{\pi}\left(R_i(t) - E[R(t)|S=S_i]\right), \\
     U_{n,2}(t) &=\sum_{s\in\mathcal{S}} \sqrt{n}(\frac{n(s)}{n} - p(s))(E[R(t)|S = s] - E[R(t)]), \\
     U_{n,3}(t)&= \sum_{s\in\mathcal{S}} \frac{D_n(s)}{\pi \sqrt{n}}E[R(t)|S = s],
\end{align*}
with $D_n(s) = \sum_{i=1}^n(A_i - \pi)I\{S_i = s\}$, $n(s) = \sum_{i=1}^nI\{S_i = s\}$ and $p(s) = P(S=s)$.
\end{lemma}
\begin{proof}
Using the fact that $E[R(t)|S=S_i] = \sum_{s\in\mathcal{S}} I\{S_i = s\}E[R(t)|S = s]$, we have
\begin{align*}
   U_{n,3}(t)= \sum_{s\in\mathcal{S}} \frac{D_n(s)}{\pi \sqrt{n}}E[R(t)|S = s] = \frac{1}{\sqrt{n}}\sum_{i=1}^n (\frac{A_i}{\pi} -1) E[R(t)|S=S_i].
\end{align*}
Using the fact that $\sum_{s\in\mathcal{S}}\sum_{i=1}^nI\{S_i = s\} = n$, $\sum_{s\in\mathcal{S}} p(s) = 1$ and $E[R(t)] = p(s)E[R(t)|S=s]$, we have
\begin{align*}
   U_{n,2}(t) &=\sum_{s\in\mathcal{S}} \sqrt{n}(\frac{n(s)}{n} - p(s))(E[R(t)|S = s] - E[R(t)]) \\
   &= \sum_{s\in\mathcal{S}}\frac{n(s)}{\sqrt{n}}(E[R(t)|S = s] - E[R(t)]) \\
   &= \frac{1}{\sqrt{n}}\sum_{i=1}^n E[R(t)|S=S_i] - \sqrt{n}E[R(t)].
\end{align*}
Hence
\begin{align*}
    & U_{n,1}(t) + U_{n,2}(t) + U_{n,3}(t) \\
    &= \frac{1}{\sqrt{n}}\sum_{i=1}^n\frac{A_i}{\pi}\left(R_i(t) - E[R(t)|S=S_i]\right) + \frac{1}{\sqrt{n}}\sum_{i=1}^n E[R(t)|S=S_i] - \sqrt{n}E[R(t)] \\
    &\qquad + \frac{1}{\sqrt{n}}\sum_{i=1}^n (\frac{A_i}{\pi} -1) E[R(t)|S=S_i] \\
    &= \frac{1}{\sqrt{n}}\sum_{i=1}^n\left(\frac{A_i}{\pi}R_i(t) - E[R(t)]\right).
\end{align*}
\end{proof}

\begin{proof}[Proof of Theorem~2.]
Under simple randomization, the results described in Theorem 2 have been proved by \cite{reid1981influence} for continuous survival functions. More generally, under simple randomization, for survival functions that may have discontinuities, the results of Theorem 2 have been proved in Section 4.3 of \cite{kosorok2008}. It remains to  prove Theorem 2 under stratified randomization and the biased-coin design, which we do below. Consider either stratified randomization or the biased coin design, and consider any $a \in \{0,1\}$.

We use $\widehat{S}_n^{(a)}$, $\widehat{\Lambda}^{(a)}, N^{(a)}, L^{(a)}$ to denote the processes $\{\widehat{S}_n^{(a)}(t): t\in [0,\tau]\}$,  $\{\widehat{\Lambda}^{(a)}(t): t\in [0,\tau]\}$, $\{N^{(a)}(t): t\in [0,\tau]\}$ and $\{L^{(a)}(t): t\in [0,\tau]\}$ respectively.
We define $\mathbb{P}_nX =\frac{1}{n}\sum_{i=1}^n X_i$ for any random variable $X$ and let $\mathbb{D}$ denote the space of real cadlag (right-continuous  with  left limit) functions with domain $[0,\tau]$ with uniform norm. For any sequence of process $\{X_n(t):t\in[0,\tau]\}$, we denote $X_n(t) = o_{p^*}(1)$ if, for any $\varepsilon>0$, $\lim_{n\rightarrow\infty}P^*(\sup_{t\in[0,\tau]}|X_n(t)|>\varepsilon) = 0$, where $P^*$ represents the outer probability measure (Section 6.2 of \citealt{kosorok2008}); that is, $o_{p*}(1)$ represents convergence to 0 in probability uniformly over $t \in [0,\tau]$.

This proof begins with showing that $\widehat{\Lambda}^{(a)}$ is asymptotically linear.
Consider the following derivation, which uses a general technique from the Appendix of \cite{Zhang2015} and Chapter 4 of \cite{kosorok2008}:
\begin{align*}
   &\widehat{\Lambda}^{(a)}(t) - \Lambda^{(a)}(t) \\
   &=  \int_{0}^t\frac{\mathbb{P}_n I\{A = a\}dN^{(a)}(t')}{\mathbb{P}_n I\{A = a\}I\{U(a) \ge t'\}} - \Lambda^{(a)}(t) \\
   &= \int_{0}^t\frac{\mathbb{P}_n I\{A = a\}dN^{(a)}(t')}{\mathbb{P}_n I\{A = a\}I\{U(a) \ge t'\}} - \int_{0}^t\frac{\mathbb{P}_n I\{A = a\}I\{U(a) \ge t'\}}{\mathbb{P}_n I\{A = a\}I\{U(a) \ge t'\}}d\Lambda^{(a)}(t')\\
   &\qquad - \int_{0}^tI\{\mathbb{P}_n I\{A = a\}I\{U(a) \ge t'\} = 0\}d\Lambda^{(a)}(t') \\
   &= \int_{0}^t\frac{\mathbb{P}_n I\{A = a\}dL^{(a)}(t')}{\mathbb{P}_n I\{A = a\}I\{U(a) \ge t'\}} - \int_{0}^tI\{\mathbb{P}_n I\{A = a\}I\{U(a) \ge t'\} = 0\}d\Lambda^{(a)}(t')\\
   &= D_{n,1}^{(a)}(t) + D_{n,2}^{(a)}(t) - D_{n,3}^{(a)}(t),
\end{align*}
where
\begin{align*}
    D_{n,1}^{(a)}(t) &= \int_{0}^t\frac{\mathbb{P}_n I\{A = a\}dL^{(a)}(t')}{\pi_aP(U(a) \ge t')}, \\
    D_{n,2}^{(a)}(t) &= \int_{0}^t\left\{\frac{1}{\mathbb{P}_n I\{A = a\}I\{U(a) \ge t'\}} -\frac{1}{\pi_aP(U(a) \ge t')} \right\} \\
    &\qquad \cdot \{\mathbb{P}_n I\{A = a\}dL^{(a)}(t')\}, \\
    D_{n,3}^{(a)}(t) &= \int_{0}^tI\{\mathbb{P}_n I\{A = a\}I\{U(a) \ge t'\} = 0\}d\Lambda^{(a)}(t').
\end{align*}

We next show $D_{n,2}^{(a)}(t) = o_{p^*}(\frac{1}{\sqrt{n}})$
using Lemma~\ref{lemma1-KM}. Since $L^{(a)}$ is a function of $U(a)$ and $\delta(a)$, then Assumption 1'(i) and stratified randomization (or the biased-coin design) imply that $(L_1^{(a)}, \dots, L_n^{(a)}) \indep (A_1,\dots, A_n) | (S_1, \dots, S_n)$. 
Defining $L^{(s,a)}$ as the conditional process of $L^{(a)}$ given $S=s$ for each $s \in \mathcal{S}$ (using Definition~\ref{def:conditional-process}), then we have that $L^{(s,a)}$ is P-Donsker by Lemmas 4.1 and Corollary 9.32 (i) of \cite{kosorok2008}, since $N^{(a)}(t),\int_0^tI\{U(a)\ge t'\}d\Lambda(t')\in \mathbb{D}$ are monotone.
Under stratified randomization or the biased-coin design, we apply our Lemma~\ref{lemma1-KM} setting $R = L^{(a)}$ and get
$\sqrt{n}\{\mathbb{P}_n I\{A = a\}L^{(a)}(t) - \pi_aE[L^{(a)}(t)]:t\in[0,\tau]\}$ weakly converges to a tight, mean 0 Gaussian process. Since Theorem~1.3.2 of \cite{fleming2011counting} implies that $L^{(a)}$ is a martingale, then
we have $E[L^{(a)}(t)] = 0$ for each $t \in [0,\tau]$ and hence the previous sentence implies that $\mathbb{P}_n I\{A = a\}L^{(a)}(t) = o_{p^*}(1)$.
By a similar argument, we have $\sqrt{n}\mathbb{P}_n \{I\{A = a\}I\{U(a) \ge t\} - \pi_aP(U(a)\ge t):t\in[0,\tau]\}$ weakly converges to a tight, mean 0 Gaussian process. 
Define $\mathbb{D}_{\phi} = \{f\in \mathbb{D}: \inf_{t\in[0,\tau]}|f(t)| > 0\}$ and $\phi: \mathbb{D}_{\phi} \mapsto \mathbb{D}$ with $\phi(g) = 1/g$. Then $\phi$ is Hadamard-differentiable,
tangentially to $\mathbb{D}$ (Page 22 of \citealp{kosorok2008}).  Since Assumption 1' (iii) implies that $P(U(a)\ge t) \in \mathbb{D}_{\phi}$, then the functional delta method (Theorem 2.8 of \citealp{kosorok2008}) implies that the process
\begin{displaymath}
 Z_n^{(a)} = \sqrt{n}\{\phi(\mathbb{P}_n I\{A = a\}I\{U(a) \ge t\}) - \phi(\pi_aP(U(a)\ge t)):t\in[0,\tau]\}
\end{displaymath}
weakly converges to a process $Z^{(a)}$.
Slutsky's theorem (Theorem 7.15 of \citealp{kosorok2008}) shows that $(Z_n^{(a)}, \mathbb{P}_n I\{A = a\}L^{(a)})$ jointly converges to $(Z^{(a)},0)$.
Since the map $ (f,g) \mapsto \{\int_0^tf(t') dg(t'): t\in[0,\tau]\}$ is continuous for $f,g\in \mathbb{D}$ and $g$ with bounded total variation, using continuous mapping theorem letting $f = Z_n^{(a)}$ and $ g = \mathbb{P}_n I\{A = a\}L^{(a)}$, we have $\sqrt{n}D_{n,2}^{(a)}(t) = \int_0^tZ_n^{(a)}(t') \mathbb{P}_n I\{A = a\}dL^{(a)}(t')$ converges to $0$ uniformly in $t$.

We then show $D_{n,3}^{(a)}(t) = o_{p^*}(\frac{1}{\sqrt{n}})$. Define $I_n(t) = I\{\mathbb{P}_n I\{A = a\}I\{U(a) \ge t\} = 0\}$. Then $I_n(t)$ is an increasing function of $t$ taking values on $\{0,1\}$, which implies that, for any $\varepsilon \in (0,1)$, $P^*(\sup_t|\sqrt{n}I_n(t)| > \varepsilon) =  P^*(\sup_t|I_n(t)| = 1) = P(I_n(\tau) = 1)$.
Since our arguments above imply that $\mathbb{P}_n I\{A = a\}I\{U(a) \ge \tau\} = \pi_a P(U(a) \ge \tau) + o_{p^*}(1)$ and Assumption 1' (iii) implies $P(U(a) \ge \tau) > 0$, then $P(I_n(\tau) = 1) = P(\mathbb{P}_n I\{A = a\}I\{U(a) \ge \tau\} = 0)\rightarrow 0$.
Hence $I_n(t) = o_{p^*}(\frac{1}{\sqrt{n}})$. Since
\begin{displaymath}
 \Lambda^{(a)}(t) \le \Lambda^{(a)}(\tau) = E\left[\int_{ 0}^{ t} \frac{dN^{(a)}(t')}{P(U(a)\ge t')}\right] \le  \frac{E[N^{(a)}(\tau)]}{P(U(a)\ge \tau)} < \infty,
\end{displaymath}
we get
\begin{displaymath}
 D_{n,3}^{(a)}(t) = \int_0^tI_n(t')d\Lambda^{(a)}(t') = \int_0^to_{p^*}(\frac{1}{\sqrt{n}})d\Lambda^{(a)}(t') \le o_{p^*}(\frac{1}{\sqrt{n}})\Lambda^{(a)}(\tau) = o_{p^*}(\frac{1}{\sqrt{n}}).
\end{displaymath}

We have shown
\begin{displaymath}
 \sqrt{n}\{\widehat{\Lambda}^{(a)}(t) - \Lambda^{(a)}(t)\} =\sqrt{n}D_{n,1}^{(a)}(t) + o_{p*}(1)= \frac{1}{\sqrt{n}}\sum_{i=1}^n\frac{I\{A_i = a\}}{\pi_a}\int_0^t\frac{d L_i^{(a)}(t')}{P(U_i(a)\ge t')} + o_{p^*}(1),
\end{displaymath}
which proves the asymptotic linearity of $\widehat{\Lambda}^{(a)}$.

We next show $\widehat{S}_n^{(a)}$ is asymptotically linear and $\sqrt{n}(\widehat{S}_n^{(a)}- S_0^{(a)})$ converges weakly to a mean 0, tight Gaussian process with covariance function $V^{(a)}(t,t')$ defined in Section C. Define $\mathbb{D}_\psi = \{f \in \mathbb{D}: \int_0^\tau|df(t')| \le M\}$ for a constant $M > 0$ large enough and product integral map $\psi: \mathbb{D}_\psi \mapsto \mathbb{D}$ with $\psi(f)(t) = \prod_{0<t'\le t} (1+d f(t))$. 
Then as shown by \cite{kosorok2008} in page 243, $\widehat{S}_n^{(a)} = \psi(-\widehat{\Lambda}^{(a)})$ and $ S_0^{(a)} = \psi(-{\Lambda}^{(a)})$. 
Lemma 12.5 of \cite{kosorok2008} implies that $\psi$ is Hadamard differentiable and we denote its derivative as $\psi'_f$, which is a continuous linear map from $\mathbb{D}_\psi$ to $\mathbb{D}$ for $f\in \mathbb{D}_\psi$. Applying the Dugundji extension theorem (Theorem 10.9 of \citealp{kosorok2008}), we can extend $\psi'_f$ to be a continuous linear map from $\mathbb{D}$ to $\mathbb{D}$ with the map on $\mathbb{D}_\psi$ unchanged. Then Theorem 20.8 of \cite{vaart_1998} implies that $\sqrt{n}\{\widehat{S}_n^{(a)}(t)- S_0^{(a)}(t)\} = \psi'_{-\Lambda}(-\sqrt{n}\{\widehat{\Lambda}^{(a)}(t) - {\Lambda}^{(a)}(t)\}) + o_{p^*}(1)$. Since $\psi'_\Lambda$ is linear and continuous, we have $ \psi'_{-\Lambda}(-\sqrt{n}\{\widehat{\Lambda}^{(a)}(t) - {\Lambda}^{(a)}(t)\}) = -\frac{1}{\sqrt{n}} \sum_{i=1}^n \frac{I\{A_i = a\}}{\pi_a}\psi'_{-\Lambda}\left(\int_0^t \frac{d L_i^{(a)}(t')}{P(U_i(a)\ge t')}\right) + \psi'_{-\Lambda}(o_{p^*}(1))$, using the asymptotic linearity of $\widehat{\Lambda}^{(a)}(t)$ proved above.
Given the formula for $\psi'_\Lambda$ in Lemma 12.5 of \cite{kosorok2008}, we have $\psi'_{-\Lambda}\left(\int_0^t \frac{d L_i^{(a)}(t')}{P(U_i(a)\ge t')}\right) = S_0^{(a)}(t)H^{(a)}_i(t)$ and $\psi'_{-\Lambda}(o_{p^*}(1)) = o_{p^*}(1)$, where $H^{(a)}_i(t)$ is defined in Section C. Hence we get
\begin{displaymath}
 \sqrt{n}\{\widehat{S}_n^{(a)}(t)- S_0^{(a)}(t)\} = -\frac{1}{\sqrt{n}} \sum_{i=1}^n \frac{I\{A_i = a\}}{\pi_a}S_0^{(a)}(t)H^{(a)}_i(t)  + o_{p^*}(1),
\end{displaymath}
which proves the asymptotic linearity claimed in Theorem. To prove the weak convergence result, we have $E[H^{(a)}(t)] = 0$ for all $t\in[0,\tau]$, using Fubini's Theorem of interchangeability of integrals (Lemma 1.2.7 of \citealp{vaart1996}).
For each $s \in \mathcal{S}$, let $H^{(a,s)}$ denote the conditional process of $H^{(a)}$ given $S=s$ (using Definition~\ref{def:conditional-process}).
Since ${H}^{(a,s)}$ is a summation of two bounded, monotone and cadlag processes,  by Lemma 4.1 and Corollary 9.32 (i) of \cite{kosorok2008}, $H^{(a,s)}$ is P-Donsker. Then we apply Lemma~\ref{lemma1-KM} and get the desired weak convergence of  $\sqrt{n}(\widehat{S}_n^{(a)}- S_0^{(a)})$ to a mean 0, tight Gaussian process with covariance function $V^{(a)}(t,t')$.

We next show $V^{(a)}(t,t) \le \widetilde{V}^{(a)}(t,t)$ for any $t \in [0,\tau]$,  where $V^{(a)}(t,t)$ is defined by equation~(\ref{def-V}) of the Supplementary Material.
Under simple randomization and Assumption 1', Theorem 1 of \cite{Zhang2015} gives that $\widetilde{V}^{(a)}(t,t) = \frac{S_0^{(a)}(t)^2}{\pi_a} Var(H^{(a)}(t))$. Then
\begin{align*}
    \widetilde{V}^{(a)}(t,t) -  V^{(a)}(t,t) = S_0^{(a)}(t)^2\frac{(1-\pi_a)}{\pi_a} E\{E[H^{(a)}(t)|S]^2\} \ge 0.
\end{align*}

Finally, we show that $V^{(a)}(t,t)$ is consistently estimated by equation (\ref{hat_Vtt}).  
For $\widehat{B}^{(a)}(t)$, direct calculation shows that 
\begin{displaymath}
\widehat{B}^{(a)}(t) = E_{n,1}^{(a)}(t) + E_{n,2}^{(a)}(t) + E_{n,3}^{(a)}(t),
\end{displaymath}
where
\begin{align*}
    E_{n,1}^{(a)}(t) &= \int_{0}^{t}\frac{d{\Lambda}^{(a)}(t') }{P(U(a)\ge t')(1 - \Delta{\Lambda}^{(a)}(t')) }, \\
    E_{n,2}^{(a)}(t) &= \int_{0}^{t}\left\{\frac{1}{\widehat{P}(U(a)\ge t')(1 - \Delta\widehat{\Lambda}^{(a)}(t'))}-\frac{1 }{P(U(a)\ge t')(1 - \Delta{\Lambda}^{(a)}(t')) }\right\}d\widehat{\Lambda}^{(a)}(t'), \\
    E_{n,3}^{(a)}(t) &= \int_{0}^{t}\frac{d\{\widehat{\Lambda}^{(a)}(t') - {\Lambda}^{(a)}(t')\} }{P(U(a)\ge t')(1 - \Delta{\Lambda}^{(a)}(t')) }.
\end{align*}

For $E_{n,1}^{(a)}(t)$, given Assumption 1' (iii), Theorem 2.6.2 of \cite{fleming2011counting} implies that $H^{(a)}$ is a martingale and $ Var(H^{(a)}(t)) = E_{n,1}^{(a)}(t)$.

For $E_{n,2}^{(a)}(t)$, we first establish $\{\Delta\widehat{\Lambda}^{(a)}(t):t\in[0,\tau]\}$ weakly converges to $\{\Delta{\Lambda}^{(a)}(t):t\in[0,\tau]\}$. Define $Q = \{I\{U(a) = t, \delta(a)=1\}:t\in[0,\tau]\}$, $Q_+ = \{I\{U(a)\ge t, \delta(a)=1\}:t\in[0,\tau]\}$ and $Q_- = \{I\{U(a)\le t, \delta(a)=1\}:t\in[0,\tau]\}$.
For each $s \in \mathcal{S}$, let $Q^{(s)}, Q_+^{(s)}, Q_-^{(s)}$ be the conditional process of $Q, Q_+, Q_-$ given $S=s$ respectively (using Definition~\ref{def:conditional-process}).
since $Q_+^{(s)}, Q_-^{(s)}$ are P-Donsker and uniformly bounded, then Corollary~9.32 of \cite{kosorok2008} implies that $Q^{(s)}$ is P-Donsker. Since Assumption 1' (iii) implies that $\inf_{t\in[0,\tau]}P(U(a)>t) > 0$, then Lemma~\ref{lemma1-KM} and continuous mapping theorem (Theorem 7.7 of \citealp{kosorok2008}) shows that $\{\Delta\widehat{\Lambda}^{(a)}(t):t\in[0,\tau]\}$ weakly converges to $\{\Delta{\Lambda}^{(a)}(t):t\in[0,\tau]\}$. Similarly, we have $\{\widehat{P}(U(a) \ge t): t\in [0,\tau]\}$ weakly converges to $\{P(U(a) \ge t): t\in [0,\tau]\}$.
Since Assumption 1' (iii) also implies that $\sup_{t\in[0,\tau]}\Delta\Lambda^{(a)}(t) < 1$, by the continuous mapping theorem, we get 
\begin{displaymath}
 \bigg|\frac{1}{\widehat{P}(U(a)\ge t')(1 - \Delta\widehat{\Lambda}^{(a)}(t'))}-\frac{1 }{P(U(a)\ge t')(1 - \Delta{\Lambda}^{(a)}(t')) }\bigg| = o_{p^*}(1).
\end{displaymath}
Then
\begin{align*}
    |E_{n,2}^{(a)}(t)| &\le \int_{0}^{t}\bigg|\frac{1}{\widehat{P}(U(a)\ge t')^2(1 - \Delta\widehat{\Lambda}^{(a)}(t'))}-\frac{1 }{P(U(a)\ge t')^2(1 - \Delta{\Lambda}^{(a)}(t')) }\bigg|d\widehat{\Lambda}^{(a)}(t') \\
    &= o_{p^*}(1) \widehat{\Lambda}^{(a)}(t).
\end{align*}
Since we have shown that $\widehat{\Lambda}^{(a)}(t) = {\Lambda}^{(a)}(t) + o_{p^*}(1)$ and ${\Lambda}^{(a)}(t)$ is uniformly bounded, we get $E_{n,2}^{(a)}(t) = o_{p^*}(1)$.

For $E_{n,3}^{(a)}(t)$, we define $T  = \{ t\in [0,\tau]: \Delta\Lambda^{(a)}(t) > 0\}$ as the set of jump points of $\Lambda^{(a)}(t)$. Since $\Lambda^{(a)}(t)$ is monotone in $t$, then $T$ is a countable set. Consider the following derivation:
\begin{align*}
    E_{n,3}^{(a)}(t) &= \int_{0}^{t}\frac{d\{\widehat{\Lambda}^{(a)}(t') - {\Lambda}^{(a)}(t')\} }{P(U(a)\ge t') } + \int_{0}^{t}\frac{ \Delta{\Lambda}^{(a)}(t')d\{\widehat{\Lambda}^{(a)}(t') - {\Lambda}^{(a)}(t')\} }{P(U(a)\ge t')(1 - \Delta{\Lambda}^{(a)}(t')) }\\
    &= \int_{0}^{t}\frac{d\{\widehat{\Lambda}^{(a)}(t') - {\Lambda}^{(a)}(t')\} }{P(U(a)\ge t') } \numberthis\label{q1}\\
    &\qquad + \sum_{t'\in T\cap[0,t]}\frac{ \Delta{\Lambda}^{(a)}(t')\{\Delta\widehat{\Lambda}^{(a)}(t') - \Delta{\Lambda}^{(a)}(t')\} }{P(U(a)\ge t')(1 - \Delta{\Lambda}^{(a)}(t')) } \numberthis\label{q2}. 
\end{align*}
For quantity~(\ref{q1}), since the map $g \mapsto \{\int_0^t\frac{dg(t')}{P(U(a)\ge t')}:t\in[0,\tau]\} $ is continuous for $g\in\mathbb{D}$ with bounded total variation, by the continuous mapping theorem, we have $\int_{0}^{t}\frac{d\{\widehat{\Lambda}^{(a)}(t') - {\Lambda}^{(a)}(t')\} }{P(U(a)\ge t') } =  o_{p^*}(1)$. For quantity~(\ref{q2}), we have
\begin{align*}
    &\sup_{t\in [0,\tau]}\bigg|\sum_{t'\in T\cap[0,t]}\frac{ \Delta{\Lambda}^{(a)}(t')\{\Delta\widehat{\Lambda}^{(a)}(t') - \Delta{\Lambda}^{(a)}(t')\} }{P(U(a)\ge t')(1 - \Delta{\Lambda}^{(a)}(t')) }\bigg| \\
     &\le \sum_{t'\in T}\frac{ \Delta{\Lambda}^{(a)}(t')|\Delta\widehat{\Lambda}^{(a)}(t') - \Delta{\Lambda}^{(a)}(t')| }{P(U(a)\ge \tau)(1 - \sup_{t'\in[0,\tau]}\Delta{\Lambda}^{(a)}(t')) } \\
    &\le \frac{ {\Lambda}^{(a)}(\tau)\sup_{t'\in[0,\tau]}|\Delta\widehat{\Lambda}^{(a)}(t') - \Delta{\Lambda}^{(a)}(t')| }{P(U(a)\ge \tau)(1 - \sup_{t'\in[0,\tau]}\Delta{\Lambda}^{(a)}(t'))}.
\end{align*}
Since ${\Lambda}^{(a)}(\tau) < \infty$, $\sup_{t'\in[0,\tau]}\Delta{\Lambda}^{(a)}(t') < 1$ and $\Delta\widehat{\Lambda}^{(a)}(t') - \Delta{\Lambda}^{(a)}(t') = o_{p^*}(1)$, then we get $(\ref{q2}) = o_{p^*}(1) $. Hence $E_{n,3}^{(a)}(t) = o_{p^*}(1)$.

Combination of the above derivations shows that  $\widehat{B}^{(a)}(t) = Var(H^{(a)}(t)) + o_{p^*}(1)$.
Following a similar proof, we can show that
\begin{displaymath}
 \frac{\sum_{i=1}^nI\{S_i = s\}\widehat{H}^{(a)}_i(t)}{\sum_{i=1}^nI\{S_i = s\}} = E[H^{(a)}(t)|S=s] + o_{p^*}(1).
\end{displaymath}
Since we have shown that $\widehat{S}_n^{(a)}(t) = S_0^{(a)}(t) + o_{p^*}(1)$, by the functional continuous mapping theorem, we have
\begin{align*}
    \widehat{V}^{(a)}(t,t)
    &= \frac{{S}_0^{(a)}(t)^2}{\pi_a}[Var(H^{(a)}(t)) - Var(E[H^{(a)}(t)|S])] + o_{p^*}(1)\\
    &= V^{(a)}(t,t) + o_{p^*}(1).
\end{align*}
\end{proof}

\section{Additional data analysis results}\label{sec:additional-analysis}
\begin{figure}[htb]
    \centering
    \includegraphics[width=\textwidth]{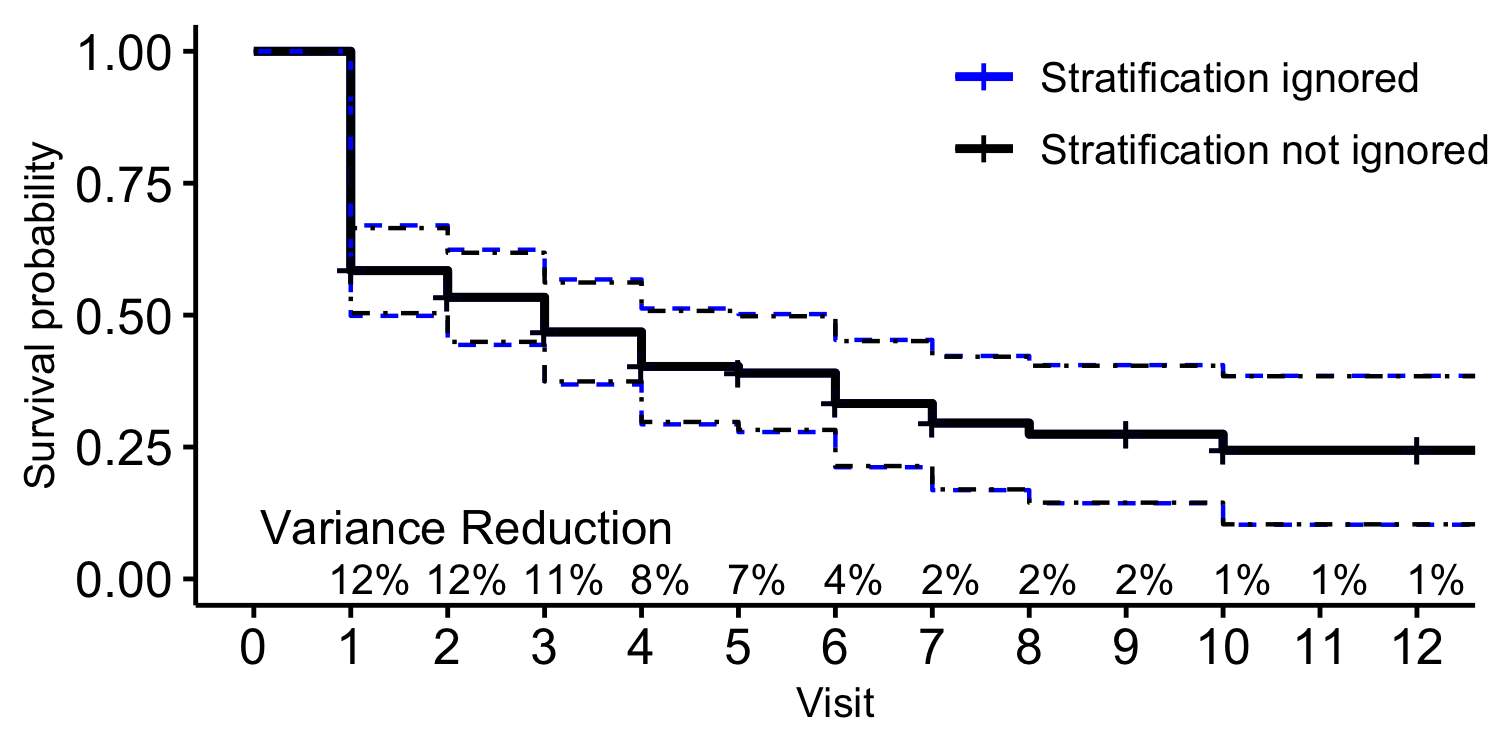}
    \caption{The K-M estimator of survival functions of NIDA-CTN-0044 control group. Dashed (dot-dashed) lines represent the estimates and confidence intervals if stratified randomization were (not) ignored in data analyses. ``Variance Reduction'' and the associated percentages represent the variance reduction by correcting the variance formula at each visit. The dashed and dot-dashed lines are very similar and almost coincide.}
    \label{fig:km}
\end{figure}

{
\bibliographystyle{biom}
\bibliography{references}
}